\documentclass[prb, 12pt, onecolumn, floatfix]{revtex4-1}

\usepackage{amssymb}
\usepackage{amsmath}
\usepackage{graphicx}
\usepackage{hyperref}
\usepackage{xcolor}
\usepackage{ulem}

\newcommand*\vect[1]{\boldsymbol{#1}}

\begin{document}

\author{Guan-Hao Peng}
\affiliation{Department of Electrophysics, National Yang Ming Chiao Tung University, Hsinchu 300, Taiwan}

\author{Chin-Jui Huang}
\affiliation{Department of Electrophysics, National Yang Ming Chiao Tung University, Hsinchu 300, Taiwan}

\author{Wen-Teng Yang}
\affiliation{Department of Electrophysics, National Yang Ming Chiao Tung University, Hsinchu 300, Taiwan}

\author{Shun-Jen Cheng}
\affiliation{Department of Electrophysics, National Yang Ming Chiao Tung University, Hsinchu 300, Taiwan}
\email{sjcheng@mail.nctu.edu.tw}

\title{Machine-learning-enabled methodology for the {\it ab-initio} simulations of sub-$\mu\text{m}$-wide nanoribbons}


\begin{abstract}

Simulation of mesoscopic nanostructures is a central challenge in condensed matter physics and device applications. First-principles methods provide accurate electronic structures but are computationally prohibitive for large systems, while empirical band theories are efficient yet limited by parameter fitting that neglects wavefunction information and often yields non-transferable parameters.
We propose a methodology that bridges these approaches, achieving first-principles-level reliability with computational efficiency through a machine-learning-enabled tight-binding framework. Our approach starts with Wannier tight-binding (WTB) parameters from small nanostructures, which serve as training data for machine-learning (ML). 
To remove the gauge freedom of Wannier functions that obscures size- and geometry-dependent parameter trends, we construct gauge-independent (GI) bases and transform the WTB model into a gauge-independent WTB (GI-WTB) model. This enables robust parameter fitting and ML prediction of parameter variations, yielding the machine-learning GI-WTB (ML-GI-WTB) model.
Applied to MoS$_{2}$ armchair-edge nanoribbons, the ML-GI-WTB model shows excellent agreement with first-principles results and enables reliable simulations of sub-$\mu$m-wide nanoribbons. 
This framework provides a scalable tool for predicting electronic properties of realistic nanostructures beyond the reach of conventional first-principles methods.

\end{abstract}

\maketitle

\section{Introduction}

Over the past few decades, \textit{ab initio} electronic structure calculations based on density functional theory (DFT), by taking advantage of the rapidly growing computing power in modern high-performance computing facilities, have played a crucial role in advancing materials science. The parameter-free DFT calculations make it possible to predict, on a first-principles basis, the ground-state properties and electronic band structures of materials, as long as the computational facilities can afford the numerical cost.
Nowadays, the DFT calculations for bulk materials or small molecular systems can be performed by using the well-established first-principles packages, e.g., VASP \cite{kresse1996efficient} and Quantum Espresso, \cite{QE-2017} at an easily affordable numerical cost. 
However, DFT calculations for nanostructures, such as nanoribbons, nanocrystals, and nanoscale devices, at the mesoscopic length scale still remain a challenging task.
This is because the translational symmetry of crystalline lattices in nanostructures is reduced or broken, and a tremendously large number of atoms need to be considered in DFT calculations. 
For instance, a realistic 50 nm-wide graphene nanoribbon consists of supercells containing around 800 carbon atoms, which is far beyond the numerical limitations of DFT.
These difficulties in the numerical implementation of DFT for nanostructures also hinder the development of nanotechnology and quantum technology based on solid-state nanostructures. \cite{chen2019monolayer}

Alternatively, simulations of nanostructures can be performed using empirical band theories, e.g., $\vect{k} \cdot \vect{p}$ and empirical tight-binding (ETB) models, constructed with a reduced number of empirically fitted parameters to reproduce band energies, and are numerically far less expensive than DFT.
However, there are several issues that limit the usefulness and validity of these empirical band theories in realistic simulations of nanostructures.

Because the set of fitting equations for the limited number of parameters in empirical band theory forms an overdetermined linear system, the parameters obtained for a specific material may vary substantially depending on the fitting procedure and algorithm. 
This non-uniqueness arises from the lack of microscopic wavefunction information, which limits the validity and applicability of such models across diverse nanostructures.
Furthermore, although empirical band theories can reproduce DFT-calculated bands accurately in certain regions of the Brillouin zone (BZ), particularly near high-symmetry points, they fail to capture the full complexity of the band structure across the entire BZ.
This limitation poses significant challenges for studies of material properties, such as exciton spectra, which require accurate band information throughout the entire BZ. \cite{peng2019distinctive,lo2021fullzone,shih2025signatures,qiu2015nonanalyticity,Deilmann_2019}

A solution to the limitations of empirical band theories is to use localized Wannier functions, transformed from a set of specified Bloch states of a material, as the basis set to construct the Wannier tight-binding (WTB) model, which is essentially equivalent to the Kohn-Sham (KS) Hamiltonian in DFT.
Based on the gauge degrees of freedom of Wannier functions, \cite{PhysRevB.56.12847} a gauge transformation matrix (a $\vect{k}$-dependent unitary matrix) can be defined to transform the KS Hamiltonian matrix from the representation of Bloch states into Wannier functions.
Because the basis transformation is unitary, the WTB parameters are determined by the DFT wavefunctions rather than the DFT band energies. In this manner, the WTB parameters are deterministic with respect to the basis set of transformed Wannier functions, and the calculated band structures in the WTB scheme almost perfectly reproduce the DFT-calculated results.
Practically, one can employ the package Wannier90 \cite{mostofi2014an} to establish the WTB model, which provides maximally-localised Wannier functions (MLWFs) as well as all Wannier function-based tight-binding parameters.
The high accuracy and physical transparency of the WTB approach make it particularly suitable for a wide range of applications beyond band structure calculations, including electron transport simulations and excited-state properties of materials. \cite{chen2019monolayer,PhysRevB.95.035430,PhysRevB.100.045411}

Although the parametrization of a DFT-based WTB scheme is deterministic, its implementation still relies on the numerical feasibility of DFT calculations, which are typically limited to bulk materials or unrealistically small nanostructures.
Attempts to use parameters from bulk or small nanostructures to construct WTB models for larger nanostructures are doomed to fail, because charge redistributes when the system geometry changes. \cite{gibertini2015emergence} This fundamental limitation reflects the non-transferability of TB parameters.
In principle, this problem could be addressed by taking Wannier function-based parameters obtained for bulk systems or small nanostructures as training data and then using interpolation or machine learning (ML) to determine the WTB parameters needed to construct DFT-based WTB models for larger nanostructures. 
However, another issue arises in the WTB scheme due to the gauge freedom in Wannier function transformations.
This gauge freedom introduces arbitrariness in the unitary transformation matrix, which leads to the non-uniqueness of the transformed Wannier functions and, consequently, makes it infeasible to directly apply ML techniques.
Because of this non-uniqueness, the Wannier function-based parameters obtained from WTB models for different nanostructures are essentially uncorrelated and typically exhibit a scattered distribution with respect to the geometric variables of the nanostructures, thereby impeding the use of interpolation or ML methods.

In this work, we present a theoretical methodology for constructing a first-principles-based WTB theory applicable to nanostructures at the sub-$\mu$m scale, beyond the computational reach of conventional DFT simulations. 
To enable data interpolation and ML, we remove the gauge freedom in Wannierization by performing a unitary transformation of the WTB Hamiltonian matrix, converting the basis of gauge-dependent Wannier functions into a specific set of atomic-orbital-like functions. 
Assuming these atomic-orbital-like functions are gauge-independent, the resulting WTB Hamiltonian matrix acquires gauge-independent parameters, which we refer to as the gauge-independent Wannier tight-binding (GI-WTB) model. 
By properly selecting these atomic-orbital-like functions, the GI-WTB parameters of different nanostructures form a consistent training dataset for ML, exhibiting clear trends with respect to the geometric variables of the nanostructures. 
This correlation enables the effective use of ML or interpolation techniques. Based on DFT-derived GI-WTB parameters for bulk and small nanostructures, ML or interpolation can then be applied to predict GI-WTB models for nanostructures of arbitrary sizes, which we denote as machine-learning gauge-independent Wannier tight-binding (ML-GI-WTB) models.

As a test nanostructured system, we apply the developed ML-GI-WTB methodology to monolayer transition-metal dichalcogenide (TMD) nanoribbons (NRs) with armchair (A) edges and calculate their electronic band structures for ribbon widths up to 200 nm. Using this approach, we perform a systematic DFT-based investigation of the width dependence of the energies and wavefunctions of TMD A-NRs over a broad range of ribbon widths, from a few nanometers to the sub-$\mu$m regime.
From the calculated band structures, we find that the energy gap rapidly converges to a constant value as the ribbon width increases. 
Analysis of the wavefunctions further reveals that low-lying conduction edge states near the band gap remain spectrally localized with only minor changes as the width increases, while states with mixed bulk-edge character will redshift toward the band gap and concentrate into a smaller spectral window as the ribbon width increase. 
High-lying conduction edge states far from the band gap are spectrally broadly distributed and overlap with bulk states in narrow NRs, while in wide NRs they concentrate into a narrow energy window, distinctly separated from the bulk states.

This paper is organized as follows.
Section II presents the fundamental theory of the WTB model, the basis transformation framework for constructing the GI-WTB model, and the parameter-fitting strategy used to develop the ML-GI-WTB model.
In Section III, we apply the theoretical framework outlined in Section II to monolayer TMD A-NRs.
This includes a statistical analysis of the hopping parameters as functions of the geometric variables of nanostructures, identification of the best-fit 2D surfaces for these datasets, and construction of the ML-GI-WTB model to predict the energy bands and wavefunctions of wide-width monolayer TMD A-NRs.
Finally, Section IV summarizes the key findings of this work.

\section{Theory}

\subsection{Kohn-Sham Equation}

The density functional theory (DFT) establishes a one-to-one correspondence between the ground state number density $\rho ( \vect{r} )$ of an $N _{e}$-electron system and its $N _{e}$-electron Hamiltonian. 
The Kohn-Sham (KS) equation implements DFT by introducing a set of KS orbitals $\{ \psi_j ( \vect{r} ) \}$, which determine the density via $\rho ( \vect{r} ) = \sum _{j = 1} ^{N _{e}} \vert \psi _{j} (\vect{r}) \vert ^{2}$. 
Thus, the interacting $N _{e}$-electron problem is recast into a single-particle KS equation for $\psi_{j} (\vect{r}) $, based on the KS Hamiltonian $H_{KS}$, which consists of the kinetic energy and $\rho$-dependent effective potentials (Hartree and exchange-correlation terms), and can be solved numerically in a self-consistent manner. 
For crystalline solids, the KS orbital is represented in Bloch form as $\psi _{n , \vect{k}} (\vect{r}) = \langle \vect{r} | \psi _{n , \vect{k}} \rangle$, labeled by the band index $n$ and Bloch wavevector $\vect{k}$, and the KS equation reads
\begin{align}\label{KS_eq}
	H_{KS} \left| \psi _{n , \vect{k}} \right\rangle = \epsilon _{n , \vect{k}} \left| \psi _{n , \vect{k}} \right\rangle,
\end{align}
where $\left| \psi _{n , \vect{k}} \right\rangle $ denotes the KS orbital state, and $\epsilon _{n , \vect{k}} $ is the eigenenergy of the KS orbital.

\subsection{Linear Combination of Atomic Orbitals Method}

In the linear combination of atomic orbitals (LCAO) method, the KS orbital (a Bloch state) is expanded as
\begin{align}\label{Bloch_state}
	\left| \psi _{n , \vect{k}} \right\rangle = \sum _{i} C _{i} ^{ (n)} (\vect{k}) \big| \phi _{i , \vect{k}} \big\rangle ,
\end{align}
in terms of the Bloch sum basis $\{\big| \phi _{i , \vect{k}} \big\rangle  \} $, defined by
\begin{align}\label{Bloch_sum_state_2}
	\big| \phi _{i , \vect{k}} \big\rangle = \frac{1}{\sqrt{N}} \sum _{\vect{R}} e^{i \vect{k} \cdot \vect{R}} \big| W _{i , \vect{R}}  \big\rangle \, ,
\end{align}
where $N$ is the total number of unit cells determined by periodic boundary conditions, $C _{i} ^{(n)} (\vect{k})$ are the expansion coefficients, and $\big\langle \vect{r} \big| W _{i , \vect{R}}  \big\rangle = W _{i} \left( \vect{r} -\vect{R} \right)$ denotes an atomic-orbital-like function localized around an atom in the unit cell at position $\vect{R}$.
In the LCAO scheme, $i \rightarrow \{ I , \alpha , s \}$ is a composite index specifying the $I$-th atom at $\vect{\tau} _{I}$ in the unit cell, the atomic orbital $\alpha$, and the electron spin $s$ for $\big| W _{i , \vect{R}}  \big\rangle$.
The Bloch sum basis of Eq.\eqref{Bloch_sum_state_2} satisfies the Bloch theorem, as does the Bloch state of Eq.\eqref{Bloch_state}. In the orthogonal tight-binding (TB) approximation, ${ \big| W_{i, \vect{R}} \big\rangle }$ is assumed to form an orthonormal basis set, which guarantees the orthonormality of the Bloch sum states.

Substituting Eq. \eqref{Bloch_state} into Eq. \eqref{KS_eq} and using the orthonormality relations, one obtains the eigenvalue equation
\begin{align}\label{TB_eigenvalue_eq}
	\sum _{j } H _{i,j} (\vect{k}) \, C_{j}^{ (n)} (\vect{k}) = \epsilon_{n , \vect{k}} \, C _{i}^{ (n)} (\vect{k}),
\end{align}
which is essentially equivalent to the KS equation but reformulated in the LCAO basis,
with $H_{i, j} (\vect{k}) \equiv \big\langle \phi _{i , \vect{k}}  \big\vert H_{KS} \big\vert \phi_{j , \vect{k}}  \big\rangle$ defining the TB Hamiltonian matrix elements. 
In TB theory, these matrix elements can be written as
\begin{align}\label{TB_matrix}
	H _{i , j} (\vect{k}) = \sum _{\vect{R}} e ^{i \vect{k} \cdot \vect{R}} \, t_{i , j} (\vect{R}),
\end{align}
where the on-site ($\vect{R} = \vect{0}$ and $\vect{\tau} _{I} = \vect{\tau} _{J}$) and hopping ($\vect{R} \neq \vect{0}$ or $\vect{\tau} _{I} \neq \vect{\tau} _{J} $) parameters are defined by $t_{i , j} (\vect{R}) = \big\langle W _{i , \vect{0}}  \big\vert H_{KS} \big\vert W _{j , \vect{R}} \big\rangle$.
Depending on the hopping distance, $ \left\vert ( \vect{R} + \vect{\tau} _{J} ) - \vect{\tau} _{I} \right\vert $, the parameters $\{t_{i , j} (\vect{R})\}$ are classified as first-, second-, third-nearest neighbors, and so forth.

In the TB theory, the eigenvalues $\epsilon _{n , \vect{k}}$ and corresponding eigenvectors $C _{i}^{ (n)} (\vect{k})$ are obtained by standard diagonalization of $H (\vect{k})$. In practice, only a finite number of Bloch sum basis states are considered to reduce the size of TB Hamiltonian matrix as long as the satisfactory convergence of numerically solved eigenenergies can be achieved.
In this work, we include five $d$-orbitals from each transition-metal atom and three $p$-orbitals from each chalcogen atom for TMD nanoribbons.
While diagonalizing Eq.~\eqref{TB_matrix} follows standard procedures, the critical challenge in TB theory lies in its universal validity and transferability, that is, how to find out the parameters that are physically reasonable and generally valid.

\subsubsection{Empirical Tight-Binding Scheme}

A common approach to determine the parameters $t _{i , j} (\vect{R})$ in Eq.~\eqref{TB_matrix} is to fit the band structure of the parametrized TB model either to experimental data or to first-principles calculations. In the former case, the number of measurable quantities, such as band gaps and effective masses, is usually limited, leading to an underdetermined system. In the latter case, the number of parameters is far smaller than the data available from continuous energy bands, resulting in an overdetermined system in which the fitted parameters depend sensitively on the chosen dataset and fitting procedure. In both cases, the fitted parameters are not unique. We refer to a TB model constructed in this way as an empirical tight-binding (ETB) model.
The limitations of ETB models stem from the absence of a proper treatment of complex wavefunctions in the fitting process, which typically considers only real-valued band energies.

\subsubsection{Wannier Tight-Binding Scheme}
\label{section_WTB}

In the Wannier tight-binding (WTB) scheme, the parameters $t _{i , j} ^{\lambda} (\vect{R}) = \big\langle W _{i , \vect{0}} ^{\lambda} \big\vert H _{KS} ^{\lambda}  \big\vert W _{j , \vect{R}} ^{\lambda} \big\rangle$ for bulk or nanostructures of a material {($\lambda$ is the system index used to distinguish different geometries, such as bulk and nanostructures)} are directly evaluated from atom-site localized states $\{ | W_{i ,\vect{R}}^\lambda \rangle \}$, known as Wannier functions. These functions are obtained from DFT-calculated Bloch states via
\begin{align}\label{Wannier_function}
	\big| W _{i , \vect{R}}^{\lambda} \big\rangle \equiv \frac{1}{\sqrt{N}} \sum _{\vect{k}} e ^{-i \vect{k} \cdot \vect{R}} \big| \phi _{i , \vect{k}}^{\lambda} \big\rangle = \frac{1}{\sqrt{N}} \sum _{\vect{k}} e ^{-i \vect{k} \cdot \vect{R}} \sum _{n = 1} ^{N _{\lambda}} U _{n , i} ^{\left( \vect{k} \right)} \left| \psi _{n , \vect{k}}^{\lambda} \right\rangle \ ,
\end{align}
where the Bloch sum basis states
\begin{align}\label{Bloch_sum_state_1}
	\big| \phi _{i , \vect{k}} ^{\lambda} \big\rangle = \sum _{n = 1} ^{N _{\lambda}} U _{n , i} ^{\left( \vect{k} \right)} \left| \psi _{n , \vect{k}} ^{\lambda} \right\rangle
\end{align}
are obtained from the Bloch states via a $\vect{k}$-dependent unitary transformation $U ^{\left( \vect{k} \right)}$. Here, $N _{\lambda}$ is the number of bands used to construct the sub-Hilbert space for system $\lambda$. Since $U ^{\left( \vect{k} \right)}$ generalizes the notion of a rotation in Euclidean space, $\big| \phi _{i , \vect{k}} ^{\lambda} \big\rangle$ is also referred to as a rotated Bloch state.

The transformation matrix $U ^{\left( \vect{k} \right)}$ is typically determined through an iterative Wannierization procedure, as implemented in the Wannier90 package. \cite{mostofi2014an}
Starting from an initial guess of $U ^{\left( \vect{k} \right)}$, often obtained by orbital projection, the procedure iteratively optimizes $U ^{\left( \vect{k} \right)}$ to minimize the spread functional of the Wannier functions. The resulting Wannier functions are known as maximally-localised Wannier functions (MLWFs). \cite{PhysRevB.56.12847,mostofi2014an} In this construction, the basis index $i$ encodes both the position and symmetry of the projecting orbital, so that Wannier functions effectively act as atomic orbitals centered on atomic sites.

Unlike ETB models, the Hamiltonian matrix in WTB model is expressed in terms of wavefunction-based parameters and is, in principle, equivalent to the KS Hamiltonian (see Fig.~\ref{fig_non_transfer}(a) for a comparison of DFT and WTB band structures).

\subsubsection{Non-Transferability of Parameters}

A general limitation of both ETB and WTB models is the non-transferability of parameters. 
A parametrization that works well for bulk materials often fails for nanostructures of the same material (see Fig.~\ref{fig_non_transfer}(b) for the TB band structure of MoS$_{2}$ nanoribbons obtained using parameters from the WTB model of 2D-bulk MoS$_{2}$ shown in (a)). 
In nanostructures, valid TB parameters must differ from those of the bulk because Bloch states are influenced not only by intrinsic material properties but also by extrinsic factors such as geometry and size. In the next section, we introduce a machine-learning strategy to predict parameter variations as the geometry of the nanostructure changes.

\subsection{Machine-Learning-Enabled Extension of WTB Theory for Nanostructures}

\label{section_parameter_fitting}

Systematically varying the nanostructure geometry allows one to derive fitting functions that explicitly capture the geometric dependence of parameters within the WTB model. These functions can then be used to predict parameters for larger-scale nanostructures by means of machine-learning (ML) enabled data-fitting procedures. In this section, we introduce a parameter-fitting scheme designed to represent the geometric dependence of WTB parameters, facilitating the calculation of electronic structures for realistically sized nanomaterials that are typically beyond the reach of direct DFT simulations.

\subsubsection{Gauge Freedom in the Transformation of Wannier Functions}
\label{machine_learing_scheme}

To reveal the geometric dependence of WTB parameters, we introduce geometric variables $g _{\ell} ^{\lambda}$ (with $\ell = 1, 2, \cdots$) to characterize the structural features of a nanostructure in a given system-$\lambda$. The WTB parameters incorporating these variables are expressed as
\begin{align}\label{hopping_parameter_g}
t _{i , j} ^{\lambda} ( \vect{R} )
= t _{I \alpha , J \beta} ^{\lambda} ( \vect{R} ) 
= t _{I \alpha , J \beta} ^{\lambda} ( \vect{R} ; \{ g _{\ell} ^{\lambda} \}) \, ,
\end{align}
where we have mapped $i \rightarrow \{ I , \alpha\}$. To keep our focus on the geometric dependence of parameters, we neglect spin-orbit coupling in this work and therefore omit the electron spin $s$ from this mapping.
The explicit definition of $g _{\ell} ^{\lambda}$ depends on the system under discussion. Specific examples will be provided later in our discussion on monolayer TMD nanoribbons.

At first glance, Eq. \eqref{hopping_parameter_g}, which explicitly depends on $g _{\ell} ^{\lambda}$, may appear adequate for capturing the geometric dependence of WTB parameters. However, the gauge freedom inherent in Wannier functions complicates this scenario. 
Since Wannier functions are constructed from Bloch sum states obtained via a unitary transformation $U ^{\left( \vect{k} \right)}$ of DFT-calculated KS orbitals (see Eqs.~\eqref{Wannier_function} and \eqref{Bloch_sum_state_1}), the resulting WTB parameters therefore depend on $U ^{\left( \vect{k} \right)}$. 
In principle, $U ^{\left( \vect{k} \right)} $ can take any unitary form, subject only to the translational invariance condition $U ^{\left( \vect{k} + \vect{G} \right)} = U ^{\left( \vect{k} \right)}$, where $\vect{G}$ is a reciprocal lattice vector. 
This gauge freedom introduces arbitrariness into the Wannier functions, ruins clear trends of WTB parameters with respect to $g _{\ell} ^{\lambda}$ when used as training data for data-fitting or ML, and ultimately hinders the development of machine-learning-enabled extensions of WTB theory for nanostructures.

To remove the influence of gauge freedom in Wannier functions, we propose the existence of gauge-independent (GI) basis set $\mathcal{S} ^{\lambda , \text{GI}} = \{ \big\vert W _{i, \vect{R}} ^{\lambda , \text{GI}} \big\rangle \} = \{ \big\vert W _{I \alpha, \vect{R}} ^{\lambda , \text{GI}} \big\rangle \}$ for each system-$\lambda$, where $\big\vert W _{I \alpha, \vect{R}} ^{\lambda , \text{GI}} \big\rangle$ serves as a atomic-orbital-like basis. The basis in $\mathcal{S} ^{\lambda , \text{GI}}$ are assumed to span the same vector space as the Wannier functions in $\mathcal{S} ^{\lambda} = \{ \big\vert W _{i, \vect{R}} ^{\lambda} \big\rangle \} = \{ \big\vert  W _{I \alpha, \vect{R}} ^{\lambda} \big\rangle \}$. Using this GI basis set, we perform a basis transformation (see the next section) on $t _{I \alpha , J \beta} ^{\lambda} (\vect{R}) = \big\langle W _{I \alpha , \vect{0}} ^{\lambda} \big\vert H _{KS} ^{\lambda}  \big\vert W _{J \beta , \vect{R}} ^{\lambda} \big\rangle$ to obtain the new WTB parameters,
\begin{align}\label{GI_hopping_parameter}
t _{I \alpha , J \beta} ^{\lambda , \text{GI}} ( \vect{R} ) = \langle W _{I \alpha , \vect{0}} ^{\lambda , \text{GI}} \vert H _{KS} ^{\lambda} \vert W _{J \beta , \vect{R}} ^{\lambda , \text{GI}} \rangle.
\end{align}
We refer to the new WTB model with parameters defined by Eq.~\eqref{GI_hopping_parameter} as the gauge-independent Wannier tight-binding (GI-WTB) model. Incorporating geometric variables as in Eq.\eqref{hopping_parameter_g}, we can express the GI-WTB parameters in Eq.~\eqref{GI_hopping_parameter} as
\begin{align}\label{GI_hopping_parameter_g}
t _{I \alpha , J \beta} ^{\lambda , \text{GI}} ( \vect{R} ) = t _{I \alpha , J \beta} ^{\lambda , \text{GI}} ( \vect{R} ; \{ g _{\ell} ^{\lambda} \} ).
\end{align}

In later discussions on nanoribbons, we will show that enforcing the constraint $\mathcal{S} ^{\lambda _{1} , \text{GI}} \in \mathcal{S} ^{\lambda _{2} , \text{GI}} \in \cdots \in \mathcal{S} ^{\lambda _{N _{\text{td}}} , \text{GI}}$, 
where $\lambda _{1}, \lambda _{2}, \cdots$ are system indices ordered by ribbon width and $N _{\text{td}}$ is the number of systems in the training dataset of our parameter-fitting scheme, ensures that the parameters in Eq.~\eqref{GI_hopping_parameter_g} acquire a well-defined geometric dependence.

Once the parameters exhibit a clear trend with respect to $g _{\ell} ^{\lambda}$, they can be fitted using the function $t _{I \alpha , J \beta} ^{\lambda, \text{ML-GI}} ( \vect{R} ; \{ g _{\ell} ^{\lambda} \} )$. By performing the replacement
\begin{align}\label{ML_GI_hopping}
	t _{I \alpha , J \beta} ^{\lambda , \text{GI}} ( \vect{R} ; \{ g _{\ell} ^{\lambda} \} ) \, \rightarrow \, t _{I \alpha , J \beta} ^{\lambda, \text{ML-GI}} ( \vect{R} ; \{ g _{\ell} ^{\lambda} \} ) , 
\end{align}
we obtain a TB model capable of predicting the electronic structure of large-scale nanostructures. We refer to the TB model with parameters, $t _{I \alpha , J \beta} ^{\lambda, \text{ML-GI}} ( \vect{R} ; \{ g _{\ell} ^{\lambda} \} )$, defined by Eq.~\eqref{ML_GI_hopping} as the machine-learning gauge-independent Wannier tight-binding (ML-GI-WTB) model.

Although the WTB parameters may not exhibit clear geometric trends as in the GI-WTB model due to gauge freedom, they can still be fitted with functions $t _{I \alpha , J \beta} ^{\lambda, \text{ML}} ( \vect{R} ; \{ g _{\ell} ^{\lambda} \} )$. The replacement $t _{I \alpha , J \beta} ^{\lambda} ( \vect{R} ; \{ g _{\ell} ^{\lambda} \} ) \, \rightarrow \, t _{I \alpha , J \beta} ^{\lambda, \text{ML}} ( \vect{R} ; \{ g _{\ell} ^{\lambda} \} )$ defines the machine-learning Wannier tight-binding (ML-WTB) model.

\subsection{Basis Transformation Theory}
\label{section}

To formulate the basis transformation theory, we have to first define the vector space under discussion. In general, we can define the vector space of system-$\lambda$ as the one spanned by the KS orbitals in the selected $N _{\lambda}$ bands, where the corresponding Bloch wavevector $\vect{k}$ are sampled on an $N$-point grid determined by periodic boundary conditions (PBCs). Within this space, the completeness relation is given by $1 ^{\lambda} = \sum _{n=1} ^{N _{\lambda}} \sum _{\vect{k}} \big| \psi _{n , \vect{k}} ^{\lambda} \big\rangle \big\langle \psi _{n ,\vect{k}} ^{\lambda} \big|$, where $1 ^{\lambda}$ is the identity operator for system-$\lambda$. Using Eqs \eqref{Bloch_sum_state_1} and \eqref{Wannier_function}, it follows that
\begin{align}\label{completeness}
1 ^{\lambda} 
= \sum _{i=1} ^{N _{\lambda}} \sum _{\vect{k}} \big| \phi _{i , \vect{k}} ^{\lambda} \big\rangle \big\langle \phi _{i , \vect{k}} ^{\lambda} \big| 
= \sum _{i=1} ^{N _{\lambda}} \sum _{\vect{R}} \big| W _{i , \vect{R}} ^{\lambda} \big\rangle \big\langle W _{i , \vect{R}} ^{\lambda} \big| ,
\end{align}
indicating that Bloch sum states and Wannier states span the same vector space as the KS orbitals.

Using Eq. \eqref{completeness}, the atomic-orbital-like basis in the set $\mathcal{S} ^{\lambda , \text{GI}}$, introduced in the previous section, can be expanded as
\begin{align}\label{basis_transform}
\big| W _{i , \vect{R}} ^{\lambda , \text{GI}} \big\rangle = \sum _{j=1} ^{N _{\lambda}} \sum _{\vect{R} ^{\prime}} S _{j , i} ^{\lambda} (\vect{R} - \vect{R} ^{\prime}) \big| W _{j , \vect{R} ^{\prime}} ^{\lambda} \big\rangle,
\end{align}
where the basis transformation matrix is defined as
\begin{align}\label{S_matrix}
S _{j , i} ^{\lambda} (\vect{R}) 
= \big\langle W _{j , \vect{0}} ^{\lambda} \big| W _{i , \vect{R}} ^{\lambda , \text{GI}} \big\rangle 
= \int _{V _{\text{SC}} ^{\lambda}} d^{3} \vect{r} \,\, W _{j , \vect{0}} ^{\lambda *} (\vect{r}) \, W _{i , \vect{R}} ^{\lambda , \text{GI}} (\vect{r}) .
\end{align}
Here, $W _{j, \vect{0}} ^{\lambda} (\vect{r}) = \big\langle \vect{r} \big| W _{j , \vect{0}} ^{\lambda} \big\rangle$ is generated by the post-processing tool Wannier90 \cite{mostofi2014an} and is localized near the atomic center in the home cell at $\vect{R} = \vect{0}$. The explicit definition of $W _{i, \vect{R}} ^{\lambda , \text{GI}} (\vect{r}) = \big\langle \vect{r} \big\vert W _{i, \vect{R}} ^{\lambda , \text{GI}} \big\rangle$ depends on the system under consideration. In later discussions on TMD nanoribbons, we will determine $W _{i, \vect{R}} ^{\lambda , \text{GI}} (\vect{r})$ by hybridizing the Wannier functions from narrow-width TMD nanoribbons and 2D-bulk TMDs.

Since the integrand in Eq.~\eqref{S_matrix} is nonzero only in the region where the two functions overlap, using a uniform $\vect{r}$-grid would waste significant computational resources in areas where the integrand vanishes. To improve efficiency, we adopt a global adaptive strategy with non-uniform $\vect{r}$-grids, which significantly reduces the number of integration points and accelerates the computation.
 
Based on Eq.~\eqref{basis_transform}, the parameters in the GI-WTB model can be evaluated as 
\begin{align}\label{similarity transformation}
	t ^{\lambda , \text{GI}} (\vect{R}) = \sum _{\vect{R} ^{\prime}} \sum _{\vect{R} ^{\prime \prime}} \, S ^{\lambda \dagger} (\vect{R} ^{\prime}) \, t ^{\lambda} (\vect{R} ^{\prime \prime}) \, S ^{\lambda} (\vect{R} - (\vect{R} ^{\prime \prime} - \vect{R} ^{\prime})) ,
\end{align}
where $t ^{\lambda , \text{GI}} _{i , j} (\vect{R}) = \big\langle W _{i , \vect{0}} ^{\lambda , \text{GI}} \big\vert H _{KS} ^{\lambda} \big\vert W _{j , \vect{R}} ^{\lambda , \text{GI}} \big\rangle$ and $t ^{\lambda} _{i , j} (\vect{R}) = \big\langle W _{i , \vect{0}} ^{\lambda} \big\vert H _{KS} ^{\lambda} \big\vert W _{j , \vect{R}} ^{\lambda} \big\rangle$.

\section{Results and Discussions}

In this work, we choose monolayer MoS$_{2}$ armchair-edge nanoribbons to demonstrate the parameter-fitting scheme proposed in Section \ref{section_parameter_fitting}.

\subsection{Monolayer MoS$_{2}$ Armchair-Edge Nanoribbons}
\label{subsec: Band structures of armchair-edge nanoribbons}

\begin{figure}
	\centering
	\includegraphics[width=\textwidth]{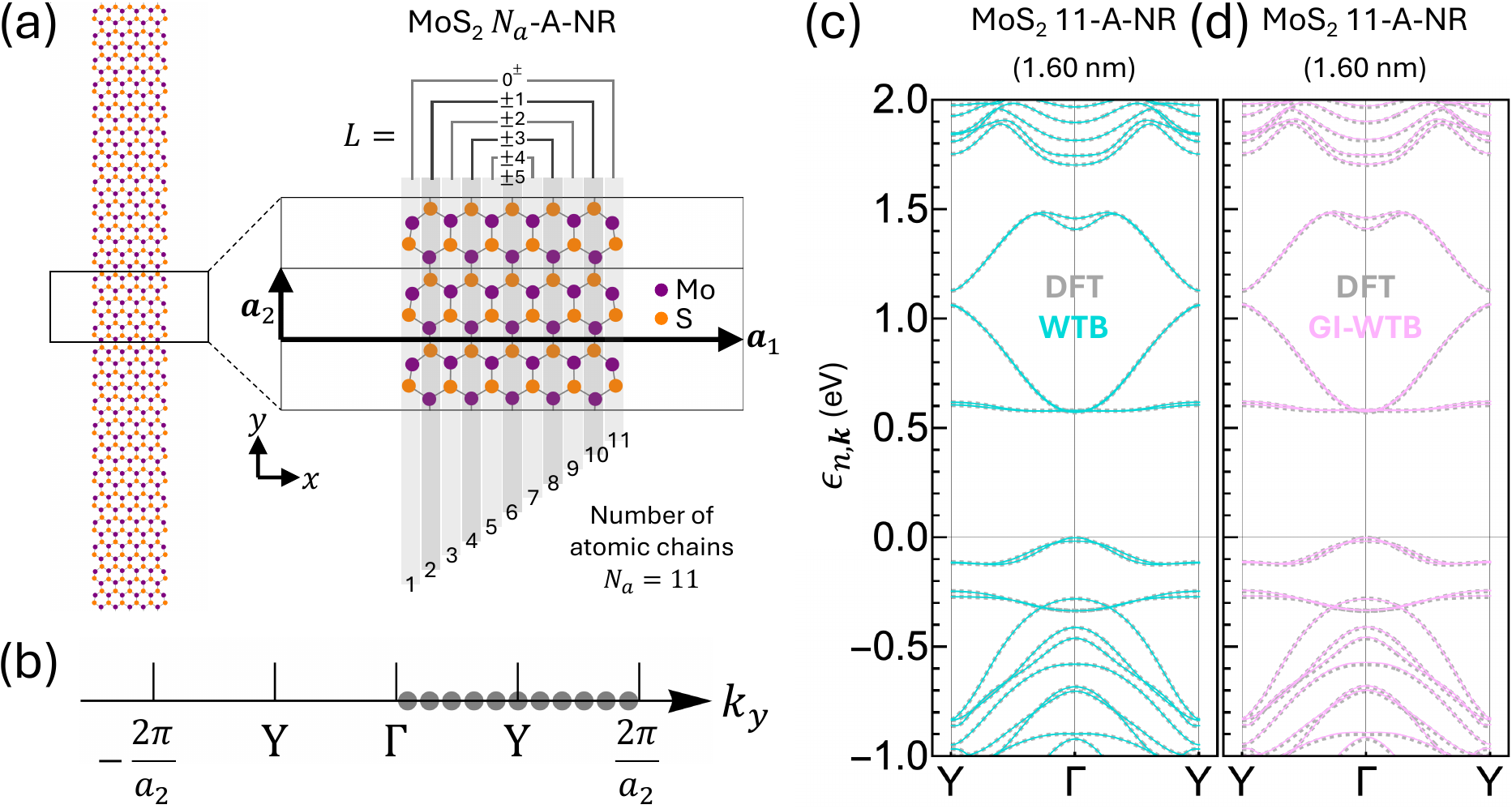}
	\caption{
	(a) Top-down view of the structure-relaxed monolayer MoS$_{2}$ armchair-edge  nanoribbon (A-NR), where the lattice translational symmetry is defined by $\vect{a} _{2} = a _{2} \hat{\vect{y}}$. Mo and S atoms are depicted in purple and orange, respectively. The integer $N _{a}$ denotes the total number of atomic chains, which characterizes the ribbon width, while the $L$-index is a geometric factor indicating the position of each atomic chain. The $L$-index equals zero at the edge and is positive or negative for chains on the left or right, respectively. Here, we refer to this ribbon as $N _{a}$-A-NR. 
	(b) The first Brillouin zone (BZ) of the $N _{a}$-A-NR, where the gray points indicate the mesh used for both the DFT calculations and Wannierization. 
	(c) Band structure of the monolayer MoS$_{2}$ 11-A-NR with a width of 1.60 nm, obtained from DFT (gray) and the Wannier tight-binding (WTB) model (cyan).
	(d) Band structure of the same nanoribbon obtained from the gauge-independent Wannier tight-binding (GI-WTB) model (pink). In all cases, the bands are aligned by shifting the valence band maximum to zero.}
	\label{fig1}
\end{figure}

Figure~\ref{fig1}(a) illustrates the atomic structure of a monolayer MoS$_{2}$ armchair-edge nanoribbon (A-NR), where the width is characterized by the number of atomic chains, $N _{a}$. For brevity, we denote this nanostructure as $N_a$-A-NR. The lattice of an $N_a$-A-NR is described by the lattice vector $\vect{R} = n _{1} \vect{a} _{1} + n _{2} \vect{a} _{2} + n _{3} \vect{a} _{3}$, where $\vect{a} _{i}$ are primitive lattice vectors, and the integers $n _{i}$ are constrained by the PBCs.
To prevent interactions between periodic images in DFT calculations, vacuum layers with thicknesses of 20 {\AA} and 16 {\AA} are introduced along $\vect{a} _{1} = a _{1} \hat{\vect{x}}$ and $\vect{a} _{3} = a _{3} \hat{\vect{z}}$, respectively. The periodicity of an $N_a$-A-NR is characterized by $\vect{a} _{2} = a _{2} \hat{\vect{y}}$, where $a _{2} = 5.52$ {\AA}.
Following the convention $\vect{a} _{i} \cdot \vect{b} _{j} = 2 \pi \delta _{ij}$, Figure~\ref{fig1}(b) presents the first Brillouin zone (BZ), defined by the primitive reciprocal lattice vector $\vect{b} _{2} = ( 2 \pi / a _{2} )\hat{\vect{y}}$.

In this work, the DFT band structures of monolayer MoS$_{2}$ $N _{a}$-A-NRs are calculated using Quantum Espresso, \cite{QE-2017} employing the generalized gradient approximation (GGA) with the Perdew-Burke-Ernzerhof (PBE) exchange-correlation functional. \cite{perdew1996generalized}
The plane-wave basis cutoff energy is set to 1088 eV, and the $\vect{k}$-mesh is sampled using a $1 \times 11 \times 1$ Monkhorst-Pack grid, represented by the gray points in Fig.~\ref{fig1}(b). 
Before structure relaxation, the atomic structures of monolayer MoS$_{2}$ $N _{a}$-A-NRs are initialized with a Mo-S bond length of 2.42 {\AA} and an out-of-plane S-S distance of 3.13 {\AA}.
Structural relaxation and self-consistent calculations are considered converged when the energy difference between consecutive iterations falls below $9.5 \times 10 ^{-4}$ eV and $9.5 \times 10 ^{-6}$ eV, respectively. Figure~\ref{fig1}(a) presents the relaxed atomic structure of a monolayer MoS$_{2}$ 11-A-NR, while the corresponding DFT band structure is shown as gray dashed lines in Fig.~\ref{fig1}(c) and (d).

Following the first-principles calculations, the DFT results are transformed into the WTB model using the post-processing tool Wannier90, \cite{mostofi2014an} which converts the Bloch states from a plane-wave representation into a Wannier representation (see Section \ref{section_WTB}). 
Figure~\ref{fig1}(c) shows the WTB band structure of the MoS$_{2}$ 11-A-NR, obtained by diagonalizing the Hamiltonian matrix in Eq.~\eqref{TB_matrix}, demonstrating excellent agreement with the DFT results.

For comparison, we also construct TB models for MoS$_{2}$ $N _{a}$-A-NRs using parameters taken from the WTB model of 2D-bulk MoS$_{2}$ (see Appendix~\ref{appendix_non_transfer}). The resulting band structure, shown in Fig.~\ref{fig_non_transfer}(b), exhibits clear discrepancies with both the DFT and WTB results. The failure of this model highlights the non-transferability of parameters, which arises because charge redistribution effects associated with edge formation are entirely neglected when bulk parameters are directly applied to NRs. \cite{gibertini2015emergence}

Although the WTB model is highly accurate, it relies on prior DFT calculations. DFT itself is limited by current high-performance computing facilities, which can handle only a few hundred atoms per unit cell. As a result, the applicability of WTB model is likewise restricted by the same computational constraints.
To overcome this bottleneck, we propose the parameter-fitting scheme introduced in Section~\ref{section_parameter_fitting}. In this approach, the WTB model is first transformed into the GI-WTB model to avoid gauge freedom in Wannier functions. 
The resulting GI-WTB parameters are then used to construct a training dataset within the geometric variable space $g _{\ell} ^{\lambda}$ of MoS$_{2}$ $N_{a}$-A-NRs. Fitting these parameters yields the ML-GI-WTB model, which enables the prediction of parameters for NRs of large width.

\subsection{The GI-WTB Model for Monolayer MoS$_{2}$ $N_{a}$-A-NRs}
\label{section_GI_WTB_of_MoS2_NR}

For monolayer MoS$_{2}$ $N _{a}$-A-NRs, we define the system index as the string $\lambda = N_{a}\text{-A-NR}$. The geometric variables introduced in Eq.~\eqref{hopping_parameter_g} are $g _{1} ^{\lambda} = N _{a}$, representing the ribbon width, and $g _{2} ^{\lambda} = L$, representing the position relative to the ribbon edge. The $L$ index for each atomic chain in system $\lambda = N _{a}\text{-A-NR}$ is illustrated in Fig.~\ref{fig1}(a). 
In this work, the edge region of a NR is defined as the atomic chains with $|L| \le 3$, while the bulk region consists of atomic chains with $|L| > 3$.

The atomic-orbital-like basis $\big\vert W _{I \alpha, \vect{R}} ^{\lambda = N _{a}\text{-A-NR} , \text{GI}} \big\rangle$ within the set $\mathcal{S} ^{\lambda = N _{a}\text{-A-NR} , \text{GI}}$ is constructed by hybridizing Wannier functions from the MoS$_{2}$ 11-A-NR and 2D-bulk MoS$_{2}$ ($\lambda = \text{2D-bulk}$).
For the edge region of $N _{a}$-A-NRs, the basis is obtained by shifting $\big\vert W _{I \alpha, \vect{R}} ^{\lambda = 11\text{-A-NR}} \big\rangle$ from the edge region of MoS$_{2}$ 11-A-NR. For the bulk region, it is obtained by shifting $\big\vert W _{I \alpha, \vect{R} = \vect{0}} ^{\lambda = \text{2D-bulk}} \big\rangle$ from the home cell of 2D-bulk MoS$_{2}$.
This procedure yields GI basis sets satisfying $\mathcal{S} ^{\lambda _{1} , \text{GI}} \in \mathcal{S} ^{\lambda _{2} , \text{GI}} \in \cdots \in \mathcal{S} ^{\lambda _{5} , \text{GI}}$, with $\lambda _{\mu} = (11 + 2 (\mu - 1))$-A-NR for $1 \le \mu \le 5$. Further details are provided in Appendix~\ref{appendix_hybridized_basis}.

With the constructed $\mathcal{S} ^{N_{a}\text{-A-NR}, \text{GI}}$, Eqs.~\eqref{S_matrix} and \eqref{similarity transformation} are used to evaluate the basis transformation matrix $S ^{N_{a}\text{-A-NR}} (\vect{R})$ and GI-WTB parameters $t ^{N _{a}\text{-A-NR} , \text{GI}} (\vect{R})$.
Substituting $t ^{N _{a}\text{-A-NR}, \text{GI}} (\vect{R})$ into Eq.~\eqref{TB_matrix} yields the Hamiltonian matrix of the GI-WTB model for MoS$_{2}$ $N _{a}$-A-NRs, which can then be diagonalized to obtain the corresponding eigenvalues and eigenvectors (see Eq.~\eqref{TB_eigenvalue_eq}).

Our definition of the edge and bulk regions in a NR, as well as the choice to construct $\mathcal{S} ^{N_{a}\text{-A-NR} , \text{GI}}$ using $\big\vert W _{I \alpha, \vect{R}} ^{11\text{-A-NR}} \big\rangle$ in the edge region of the MoS$_{2}$ 11-A-NR and $\big\vert W _{I \alpha, \vect{R} = \vect{0}} ^{\text{2D-bulk}} \big\rangle$ in the home cell of 2D-bulk MoS$_{2}$, is guided by both physical intuition and numerical validation.

From a physical perspective, for the edge states of NRs, it is reasonable to assume that the charge distribution extends only a limited distance from the edges and becomes stable once the ribbon is sufficiently wide. For ribbons with $N _{a} \ge 11$, the edge-state charge distribution is expected to remain stable and localized within the region $L \le 3$. Likewise, for bulk states, the charge distribution is expected to localize near the ribbon center as the width increases, and can be effectively described by the Wannier functions of the 2D-bulk system, which are localized within the region $L > 3$.

Numerical tests validate our assumption. By diagonalizing the Hamiltonian matrix of the GI-WTB model constructed using $\mathcal{S} ^{N_{a}\text{-A-NR} , \text{GI}}$, the resulting band structures, shown as cyan lines in Fig.~\ref{fig1}(d), exhibit excellent agreement with the DFT results. Further consistent results between the GI-WTB model and DFT for $N _{a} > 11$ are provided in Appendix \ref{appendix_GI_WTB_bands}. 
Although we have not analytically proven that $\mathcal{S} ^{N_{a}\text{-A-NR} , \text{GI}}$ spans the same space as $\mathcal{S} ^{N_{a}\text{-A-NR}}$, the numerical results clearly demonstrate its suitability and reliability. 

As a technical remark, the iteration steps in the Wannierization process are crucial to our basis transformation theory. In Wannier90, \cite{mostofi2014an} the center and profile of Wannier functions evolve during the iteration process to minimize the spread functional. Since the overlap between two Wannier functions can change significantly due to minor adjustments in their profiles and centers, numerous iterations can introduce unforeseen changes in the basis transformation matrix defined in Eq.~\eqref{S_matrix}, leading to instability.
To address this, we adopt a \textit{one-shot} Wannierization procedure for both $W _{j, \vect{0}} ^{N _{a}\text{-A-NR}} (\vect{r})$ and the Wannier functions used to construct $W _{i , \vect{R}} ^{N _{a}\text{-A-NR} , \text{GI}} (\vect{r})$ for evaluating $S _{j , i} ^{N _{a}\text{-A-NR}} (\vect{R})$ in Eq.~\eqref{S_matrix}. In this approach, the matrix $U ^{(\vect{k})}$ is determined in a single step using the orbital projection method. \cite{PhysRevB.56.12847}
The resulting Wannier functions closely preserve the intended profiles and centers specified by the projection orbitals. By shifting these \textit{well-behaved} Wannier functions to the atomic sites of the $N _{a}$-A-NRs, our numerical tests confirm that the basis transformation results are stable and reliable.

\subsection{Parameter-Fitting for Monolayer MoS$_{2}$ $N_{a}$-A-NRs}

\begin{figure}
	\centering
	\includegraphics[width=\textwidth]{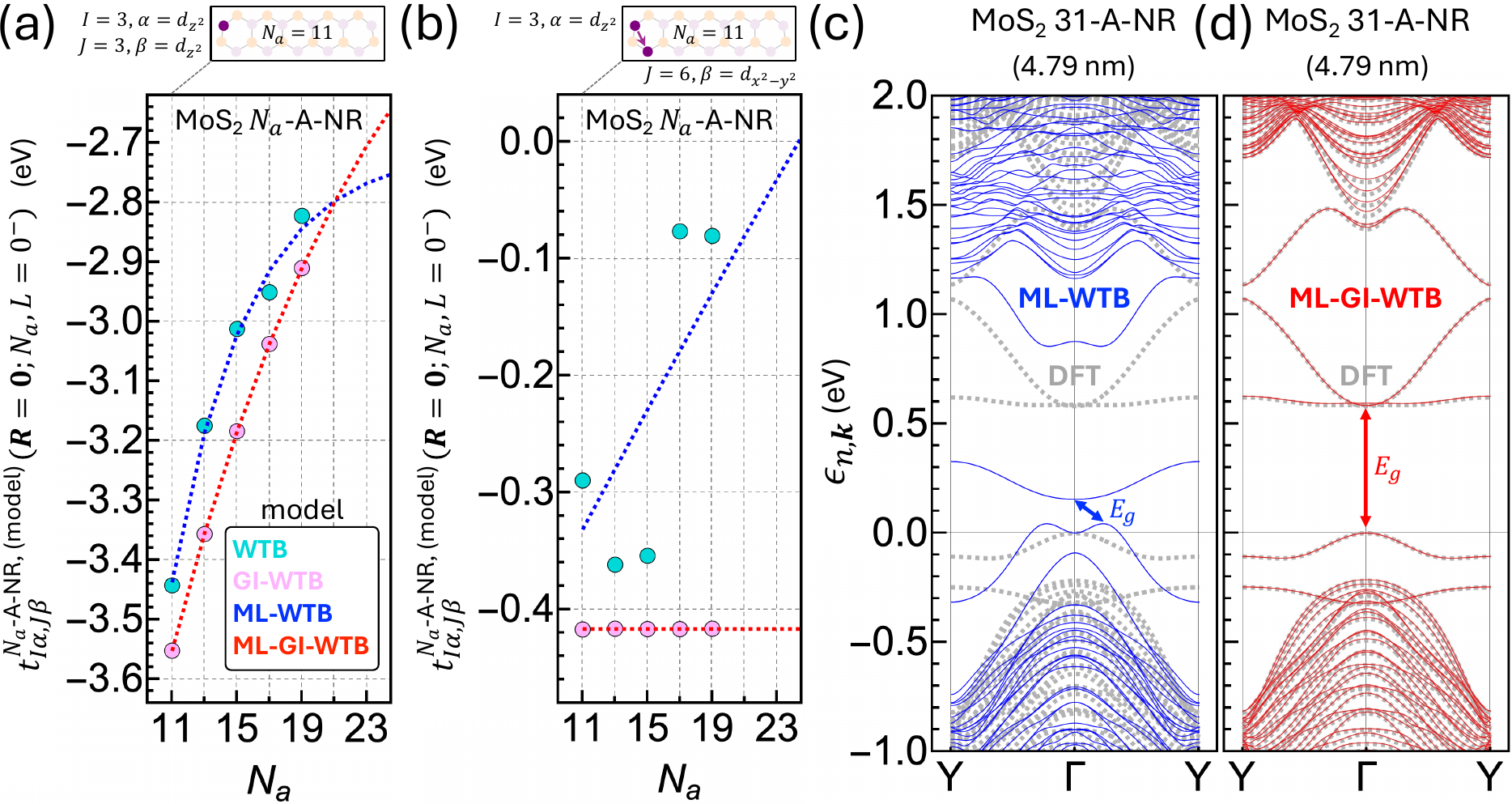}
	\caption{
	(a) $N _{a}$ dependence of the on-site energy for the $d _{z ^{2}}$-orbital at the atom indicated in the top schematic (illustrated using 11-A-NR). 
	(b) $N _{a}$ dependence of the hopping parameter between the $d _{z ^{2}}$- and $d _{x ^{2} - y ^{2}}$-orbitals, as denoted by the arrow in the top schematic (again shown using 11-A-NR). 
	The cyan and pink data points are obtained from the WTB and GI-WTB models, respectively. 
	The blue dashed line for the machine-learning WTB (ML-WTB) model is fitted to the cyan data points, while the red dashed line for the machine-learning GI-WTB (ML-GI-WTB) model is fitted to the pink data points. 
	(c) Band structure of monolayer MoS$_{2}$ 31-A-NR from the ML-WTB model (blue) compared with DFT (gray). 
	(d) Band structure of monolayer MoS$_{2}$ 31-A-NR from the ML-GI-WTB model (red) compared with DFT (gray). 
	Double-headed arrows indicate the energy band gap, $E _{g}$.}
	\label{fig2}
\end{figure}

To demonstrate the advantages of the GI-WTB model, we analyze parameters in the geometric variable space defined by $g _{1} ^{\lambda} = N _{a}$ and $g _{2} ^{\lambda} = L$. 
In Fig.~\ref{fig2}(a), we present parameters from the WTB model, $t _{I \alpha, J \beta} ^{N _{a}\text{-A-NR}} (\vect{R} ; N _{a}, L)$, and from the GI-WTB model, $t _{I \alpha, J \beta} ^{N _{a}\text{-A-NR}, \text{GI}} (\vect{R} ; N _{a}, L)$, for $I = J = 3$, $\alpha = \beta = d _{z ^{2}}$, $\vect{R} = \vect{0}$, $L = 0$, and $N _{a} = 11, 13, 15, 17, 19$. These correspond to the on-site energies in the TB model. The schematic at the top of Fig.~\ref{fig2}(a) illustrates the orbital center associated with the on-site energy for the 11-A-NR.

To examine how the gauge freedom of Wannier functions affects the geometric dependence of parameters, we generated the WTB model by performing Wannierization with different iteration steps. The numbers of iterations were 20000 for $\lambda _{1} = 11$-A-NR, 200 for $\lambda _{2} = 13$-A-NR, 300 for $\lambda _{3} = 15$-A-NR, and 10000 for both $\lambda _{4} = 17$-A-NR and $\lambda _{5} = 19$-A-NR. In contrast, the GI-WTB model was constructed by transforming the basis sets of the corresponding one-shot WTB models into the GI basis sets introduced and validated in the previous section.

From Fig.~\ref{fig2}(a), the dataset for the GI-WTB model exhibits a more systematic and consistent trend compared to that of the WTB model. This contrast indicates that the gauge freedom inherent in Wannier functions can obscure the geometric dependence of parameters. By removing gauge effects, the GI-WTB model provides a well-defined and robust geometric dependence.
A more illustrative comparison is shown in Fig.~\ref{fig2}(b), which presents the hopping parameters from the WTB and GI-WTB models for the case with $I = 3$, $\alpha = d _{z ^{2}}$, $J = 6$, $\beta = d _{x ^{2} - y ^{2}}$, $\vect{R} = \vect{0}$, $L = 0$, and $N _{a} = 11, 13, 15, 17, 19$. The purple arrow in the schematic at the top of Fig.~\ref{fig2}(b) indicates the tunneling vector of the two involved orbitals. The WTB model shows a scattered geometric dependence, reflecting the influence of gauge freedom. In contrast, the GI-WTB model produces a smooth and consistent trend, demonstrating its gauge-independent nature.

In our fitting procedure, the GI-WTB parameters in Eq.~\eqref{GI_hopping_parameter_g} are first categorized by the orbital indices ($\alpha$ and $\beta$) and tunneling vector $\vect{d} \equiv ( \vect{\tau} _{J} + \vect{R} ) - \vect{\tau} _{I} $. 
For example, Figure~\ref{fig2}(a) belongs to the category $\{ \alpha = d _{z ^{2}}, \beta = d _{z ^{2}}, \vect{d} = \vect{0} \}$, while Fig.~\ref{fig2}(b) belongs to $\{ \alpha = d _{z ^{2}}, \beta = d _{x ^{2} - y ^{2}}, \vect{d} = \vect{d} _{0} \}$, with $\vect{d} _{0}$ indicated by the purple arrow in the top-schematic of Fig.~\ref{fig2}(b). 
Within each category, parameters are organized into a training dataset on the two-dimensional geometric variable space spanned by $g _{1} ^{\lambda} = N _{a}$ and $g _{2} ^{\lambda} = L$. 
For instance, Figure~\ref{fig2}(a) shows only the subset at $L = 0 ^{-}$. To build the complete dataset, the same plotting procedure as in Fig.~\ref{fig2}(a) is repeated for all other $L$ values ($0^{+}, \pm1, \pm2, \cdots$). By combining these plots, we obtain the full training dataset over the $N _{a}$-$L$ plane for the category $\{ \alpha = d _{z^{2}}, \beta = d _{z ^{2}}, \vect{d} = \vect{0} \}$. 
A similar procedure is applied to Fig.~\ref{fig2}(b) and to other on-site and hopping parameters. 
The fitting procedure is likewise applied to the WTB parameters in Eq.~\eqref{hopping_parameter_g}.

After constructing the training dataset, we fit the parameters using the functions $t _{I \alpha , J \beta} ^{\lambda , \text{ML-GI}} (\vect{R} ; \{ g _{\ell} ^{\lambda} \})$ and $t _{I \alpha , J \beta} ^{\lambda , \text{ML}} (\vect{R} ; \{ g _{\ell} ^{\lambda} \})$ introduced in Section~\ref{machine_learing_scheme}.
For datasets symmetric with respect to the $L$-axis in geometric variable space, we assume $t _{I \alpha , J \beta} ^{\lambda , \text{ML-GI}} (\vect{R} ; N _{a} , L ) = \delta _{1} + \delta _{2} \exp(-\delta _{4} \vert L \vert) + \delta _{3} \exp(- \delta _{5} N _{a})$ and $t _{I \alpha , J \beta} ^{\lambda , \text{ML}} (\vect{R} ; N _{a} , L ) = \gamma _{1} + \gamma _{2} \exp(-\gamma _{4} \vert L \vert) + \gamma _{3} \exp(- \gamma _{5}  N _{a})$, where $\delta _{\mu}$ and $\gamma _{\mu}$ are fitting parameters. 
For datasets anti-symmetric with respect to the $L$-axis, we assume $t _{I \alpha , J \beta} ^{\lambda , \text{ML-GI}} (\vect{R} ; N _{a} , L ) = \text{sgn}(L) \,  \big[ \delta _{1} + \delta _{2} \exp(-\delta _{4} \vert L \vert) + \delta _{3} \exp(- \delta _{5} N _{a}) \big]$ and $t _{I \alpha , J \beta} ^{\lambda , \text{ML}} (\vect{R} ; N _{a} , L ) = \text{sgn}(L) \,  \left[ \gamma _{1} + \gamma _{2} \exp(-\gamma _{4} \vert L \vert) + \gamma _{3} \exp(- \gamma _{5}  N _{a}) \right]$, where sgn$()$ denotes the sign function. 
In MoS$_{2}$ $N _{a}$-A-NRs, all datasets fall into either the symmetric or antisymmetric category and can be fitted using these functional forms. 
To ensure the expected decay behavior, we require $\gamma _{4} > 0$, $\gamma _{5} > 0$, $\delta _{4} > 0$, and $\delta _{5} > 0$ in this study.

To determine the fitting parameters $\delta _{\mu}$ and $\gamma _{\mu}$ in the assumed fitting function, we apply the least-squares method by minimizing the residual functions
\begin{align}\label{residual_function_GI}
 	\Delta _{\alpha , \beta , \vect{d}} (\vect{\delta})
 	= \sum _{N _{a}} \sum _{L} \left| t _{I \alpha , J \beta} ^{N _{a}\text{-A-NR} , \text{ML-GI}} (\vect{R} ; N _{a} , L ) - t _{I \alpha, J \beta} ^{N _{a}\text{-A-NR} , \text{GI}} (\vect{R} ; N _{a}, L) \right| ^{2}
\end{align}
and 
\begin{align}\label{residual_function}
 	\Gamma _{\alpha , \beta , \vect{d}} (\vect{\gamma})
 	= \sum _{N _{a}} \sum _{L} \left| t _{I \alpha , J \beta} ^{N _{a}\text{-A-NR} , \text{ML}} (\vect{R} ; N _{a} , L ) - t _{I \alpha, J \beta} ^{N _{a}\text{-A-NR}} (\vect{R} ; N _{a}, L) \right| ^{2} \, ,
\end{align}
where $\vect{\delta} = \sum _{\mu = 1} ^{5} \hat{\vect{e}} _{\mu} \delta _{\mu}$, and $\vect{\gamma} = \sum _{\mu = 1} ^{5} \hat{\vect{e}} _{\mu} \gamma _{\mu}$. 
At first glance, the indices $\{ I , J , \vect{R} \}$ in Eqs.\eqref{residual_function_GI} and \eqref{residual_function} may appear undetermined. In fact, they are constrained by the condition $\vect{d} = (\vect{\tau} _{J} + \vect{R}) -\vect{\tau} _{I}$, which defines the training dataset. Each $\{ I , J , \vect{R} \}$ satisfying this condition  corresponds uniquely to a coordinate $(N _{a} , L)$ in the geometric variable space. Therefore, when summing over all $(N _{a} , L)$ points in the training dataset, all valid $\{ I , J , \vect{R} \}$ are automatically included, ensuring that no indices remain ambiguous.

To secure the correct asymptotic behavior, we impose boundary conditions during the optimization. These conditions require the fitting functions to converge to the WTB parameters of 2D-bulk MoS$_{2}$,
$t _{I \alpha , J \beta} ^{N _{a}\text{-A-NR} , \text{ML-GI}} (\vect{R} ; N _{a} \rightarrow \infty , L \rightarrow \pm \infty) = t _{I \alpha , J \beta} ^{N _{a}\text{-A-NR} , \text{ML}} (\vect{R} ; N _{a} \rightarrow \infty , L \rightarrow \pm \infty) = t _{I \alpha , J \beta} ^{\text{2D-bulk}} (\vect{R})$, 
during the minimization of Eqs.~\eqref{residual_function_GI} and \eqref{residual_function}.
This condition fixes the parameters $\delta _{1}$ and $\gamma _{1}$ in the fitting functions.
By applying the replacement in Eq.~\eqref{ML_GI_hopping} to the Hamiltonian matrix in Eq.~\eqref{TB_matrix}, we can obtain the ML-GI-WTB model. The same procedure is also used to construct the ML-WTB model (see Section~\ref{machine_learing_scheme}).

In Fig.~\ref{fig2}(a) and (b), the ML-WTB parameters, $t _{I \alpha , J \beta} ^{N _{a}\text{-A-NR} , \text{ML}} ( \vect{R} ; N _{a} , L)$, and the ML-GI-WTB parameters, $t _{I \alpha , J \beta} ^{N _{a}\text{-A-NR} , \text{ML-GI}} ( \vect{R} ; N _{a} , L)$, are shown as blue and red dashed lines, respectively. A clear discrepancy is observed between the WTB and ML-WTB results for the on-site energies when $N _{a} > 15$, as shown in Fig.~\ref{fig2}(a), with further deviations evident in the hopping terms shown in Fig.~\ref{fig2}(b). In contrast, the GI-WTB data points exhibit excellent agreement with the corresponding fitting curves in the ML-GI-WTB model for both on-site energies and hoppings, highlighting the robustness of the proposed GI-WTB model.

Using the ML-WTB and ML-GI-WTB models, we can predict the parameters for MoS$_{2}$ $N _{a}$-A-NRs with $N _{a} > 19$, which lie beyond the range of the training dataset (see Fig.~\ref{fig2}(a) and (b)). In this regime, the predictions from the ML-WTB model are expected to be inaccurate, while those from the ML-GI-WTB model remain reliable. Based on the predicted parameters, the corresponding TB Hamiltonian matrices are constructed using Eq.\eqref{TB_matrix}.
Figures~\ref{fig2}(c) and (d) compare the DFT band structure (gray dashed lines) of monolayer MoS$_{2}$ 31-A-NR with the results obtained from the ML-WTB model (blue lines) and the ML-GI-WTB model (red lines), respectively. The ML-GI-WTB model reproduces the DFT bands with excellent accuracy, whereas the ML-WTB model produces significant deviations.

\begin{figure}
	\centering
	\includegraphics[width=\textwidth]{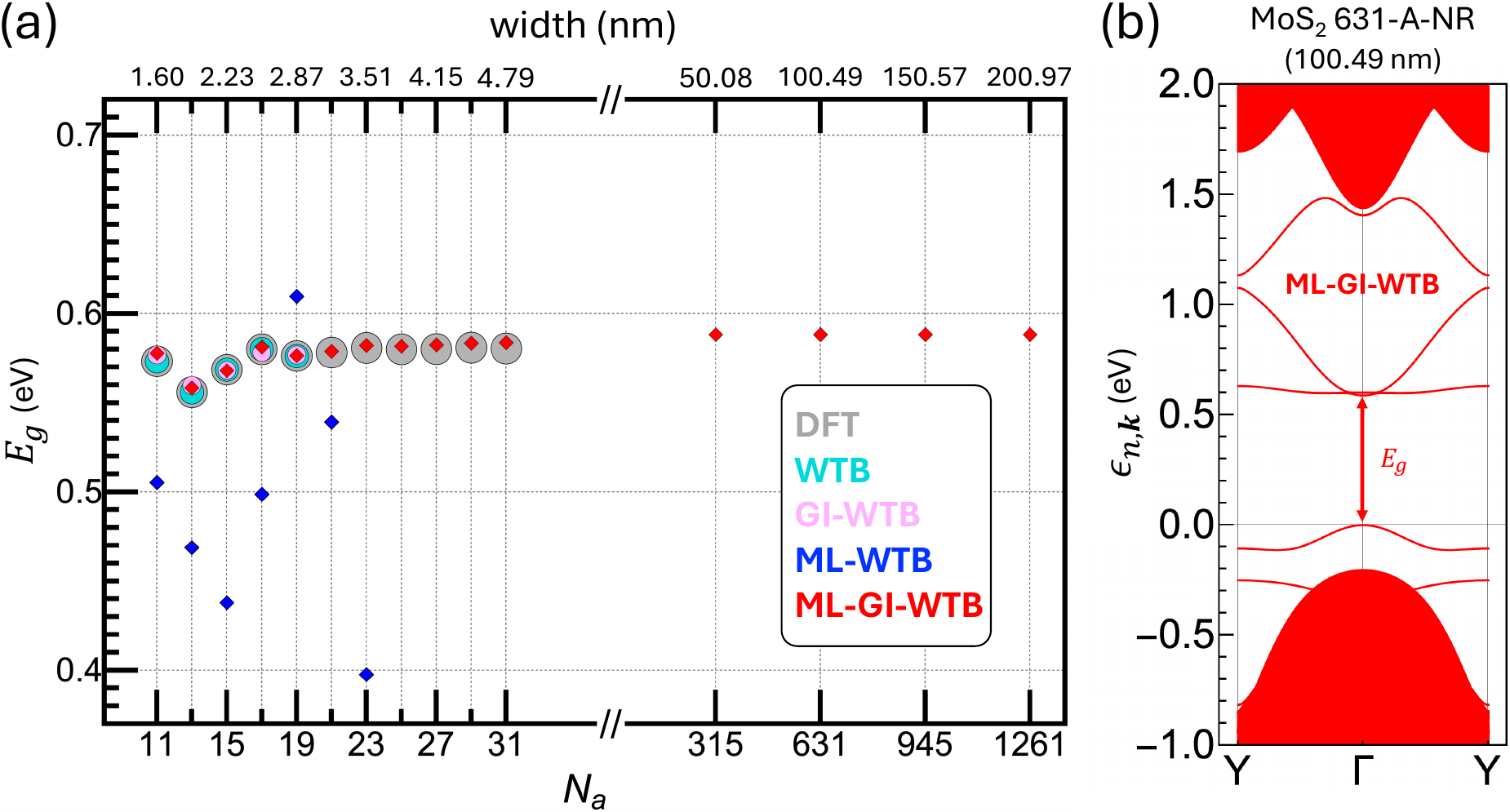}
	\caption{
	(a) Dependence of the energy gap $E _{g}$ on $N _{a}$, ranging from $N _{a} = 11$ to $N _{a} = 1261$. The upper axis shows the corresponding widths of the monolayer MoS$_{2}$ $N _{a}$-A-NR. Data points are color-coded according to the model, with gray for DFT, cyan for WTB, pink for GI-WTB, blue for ML-WTB, and red for ML-GI-WTB. 
	(b) Band structure of the sub-$\mu$m-wide monolayer MoS$_{2}$ 631-A-NR computed using the ML-GI-WTB model.}
	\label{fig3}
\end{figure}

To further demonstrate the effectiveness of our approach, Figure~\ref{fig3}(a) presents the energy band gap $E _{g}$ for monolayer MoS$_{2}$ $N _{a}$-A-NRs, starting from $N _{a} = 11$ (1.60 nm) to $N _{a} = 1261$ (200.97 nm). As a reference, we compute the DFT results for $E _{g}$ up to $N _{a} = 31$. Within this range, the ML-GI-WTB model exhibits excellent agreement with the DFT results. In contrast, the ML-WTB model yields disorganized and significantly deviated $E _{g}$ values, reflecting the limitations introduced by the gauge freedom in Wannier functions for parameter fitting or ML purpose. 
As seen in Fig.~\ref{fig3}(a), the band gap saturates to a constant value as ribbon width increases. For illustration, Fig.~\ref{fig3}(b) presents the band structure of a MoS$_{2}$ 631-A-NR (100.49 nm wide), where the spectrum exhibits nearly continuous valence and conduction bands at higher energies.

\subsection{Ribbon-Width Dependence of State Probability Distributions}

\begin{figure}
	\centering
	\includegraphics[width=\textwidth]{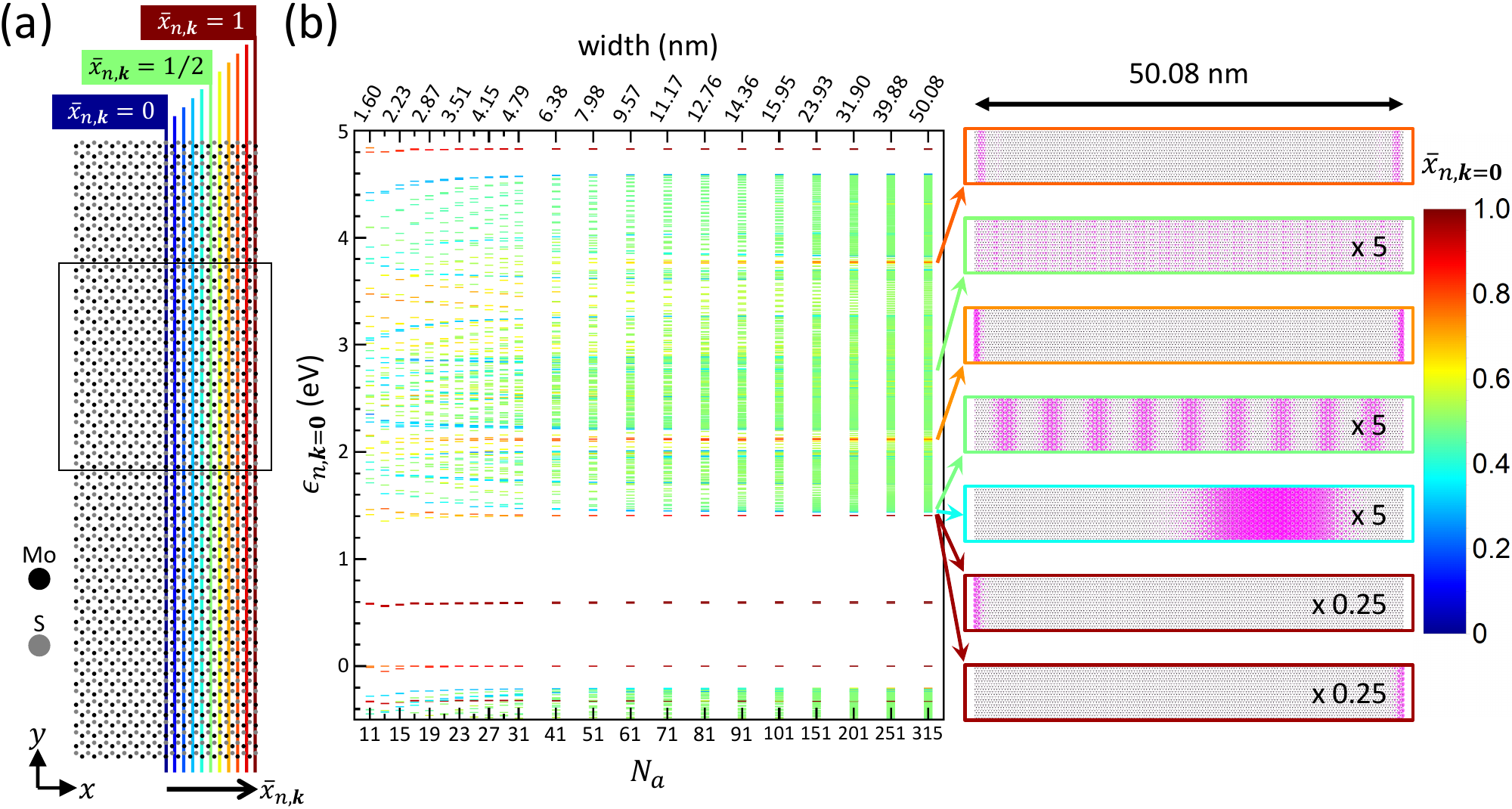}
	\caption{
	(a) Schematic illustration of the relative average position $\bar{x} _{n \vect{k}}$ for the Bloch state $| \psi _{n , \vect{k}} \rangle$ (see definition in Eq.~\eqref{x_bar}) in monolayer MoS$_{2}$ $N_{a}$-A-NR. The deep blue line marks $\bar{x} _{n \vect{k}} = 0$ at the ribbon center. The deep red line marks $\bar{x} _{n \vect{k}} = 1$ at the ribbon edge. The light green line marks $\bar{x} _{n \vect{k}} = 1/2$ at the midpoint between the ribbon center and the edge. Other values of $\bar{x} _{n \vect{k}}$ between $0$ and $1$ are marked by colored lines as indicated. Mo atoms are shown in black and S atoms are shown in gray to avoid confusion from overuse of colors. 
	(b) Energy levels at the $\Gamma$-point for MoS$_{2}$ $N_{a}$-A-NRs of different widths, where the $\bar{x} _{n \vect{k}}$ of each Bloch state is color-coded according to the scale on the right and demonstrated in (a). For the 50.08 nm wide NR, several states are selected as representative examples of the probability distributions (indicated by color-coded arrows). The radius of the magenta circles represents the probability at each atomic site. Numbers at the upper right of each probability plot indicate the applied scaling factors.
	}
	\label{fig4}
\end{figure}

In NRs, identifying the spatial probability distribution of eigenstates, including bulk and edge states, is essential for practical applications. The proposed ML-GI-WTB model is a powerful tool for this purpose, as it provides direct access to wavefunction information in wide-width NRs, well beyond the reach of conventional DFT calculations. In this section, we will study the ribbon-width dependence of state probability distributions through defining the relative average position of each energy eigenstate.

In the TB model, the composition weight of the Bloch sum state $| \phi _{I \alpha , \vect{k}} \rangle$ in the band state $| \psi _{n , \vect{k}} \rangle$ is given by the norm squared of the linear combination coefficient $\big| C _{I \alpha} ^{(n)} (\vect{k}) \big| ^{2}$ (see Eq.~\eqref{Bloch_state}). 
Since $| \phi _{I \alpha , \vect{k}} \rangle$ is periodically localized at $\vect{\tau} _{I}$ within each unit cell through the localized basis functions $| W _{I \alpha , \vect{R}} \rangle$ (see Eq.~\eqref{Bloch_sum_state_2}), summing $\big| C _{I \alpha} ^{(n)} (\vect{k}) \big| ^{2}$ over different orbital indices $\alpha$ but fixing the atomic position index $I$ may be interpreted (although not strictly) as the probability of finding the quasi-particle at $\vect{\tau} _{I}$ within each unit cell.

To facilitate the following analysis, we set the origin of the $x$-axis at the ribbon center. Accordingly, the probability of finding the quasi-particle at the atomic chain indexed by $L$ in a NR (see Fig.~\ref{fig1}(a)) can be written as
\begin{align}\label{prob_at_L}
P _{n , \vect{k}} ^{L} 
= \sum _{\substack{I \\ \tau _{I , x} = x _{L}}} \sum _{\alpha \in \mathcal{A} _{\Theta (I)}} \big| C _{I \alpha} ^{(n)} (\vect{k}) \big| ^{2} ,
\end{align}
where $x _{L} = \pm \left( \frac{N _{a} - 1}{2} \right) x _{0} \mp |L| \, x _{0}$ is the $x$-position of the atomic chain for $L = \pm |L|$, $x _{0}$ is the spacing between atomic chains, and $\Theta ( I ) = \delta _{\text{mod} (I , 3) , 0} + 1$ is the atomic-species function. The orbital set $\mathcal{A} _{1} = \{ p _{z} , p _{x} , p _{y} \}$ corresponds to the $p$-orbitals of the chalcogen atoms, and the orbital set $\mathcal{A} _{2} = \{ d _{z ^{2}} , d _{xz} , d _{yz} , d _{x ^{2} - y ^{2}} , d _{xy} \}$ corresponds to the $d$-orbitals of the transition-metal atoms (see Appendix~\ref{appendix_hybridized_basis} for details).

We characterize the probability distribution of a band state by defining
\begin{align}\label{x_bar}
	\bar{x} _{n , \vect{k}} = \frac{\sum _{L} P _{n , \vect{k}} ^{L} \, | x _{L} |}{w/2} ,
\end{align}
where $w = (N _{a} - 1) x _{0}$ is the ribbon width. In Eq.\eqref{x_bar}, the numerator gives the average position of the Bloch state, and dividing by $w/2$ yields its relative average position within the NR.
Figure~\ref{fig4}(a) illustrates the interpretation of $\bar{x} _{n , \vect{k}}$. Bulk states may fall in the range $0 \le \bar{x} _{n , \vect{k}} \le 0.5$, edge states may fall in the range $0.5 \le \bar{x} _{n , \vect{k}} \le 1$, and bulk-edge mixed states may appear around $\bar{x} _{n , \vect{k}} \approx 0.5$.

Figure~\ref{fig4}(b) presents the energy spectra of band states at the $\Gamma$-point for MoS$_{2}$ $N _{a}$-A-NRs of different widths, where the relative average position $\bar{x} _{n , \vect{k}}$ is color-coded according to the scheme illustrated in Fig.~\ref{fig4}(a). The states near $0$ eV and $0.6$ eV correspond to the valence band maximum and conduction band minimum, respectively, which remain stable with varying ribbon width, consistent with the band gap $E _{g}$ behavior shown in Fig.~\ref{fig3}(a).
Low-lying conduction edge states (yellow to red) close to the band gap remain spectrally localized and show only minor shifts as the width increases. 
States with mixed bulk-edge character (cyan) redshift toward the band gap and concentrate into a smaller spectral window as the ribbon width increases.
High-lying conduction edge states (yellow to orange) far above the gap are broadly distributed and overlap with bulk states (green) in narrow ribbons, but in wide ribbons they converge into a narrow energy window, becoming distinctly separated from the bulk spectrum.

To confirm that $\bar{x} _{n , \vect{k}}$ provides a reliable measure of the spatial distribution of band states, we also plot the probability distribution from the first sum (the sum over $\alpha$) in Eq.~\eqref{prob_at_L} for the 50.08 nm MoS$_{2}$ 315-A-NR, as indicated by the color-coded arrows in Fig.~\ref{fig4}(b). 
The results reproduce the same trends captured by $\bar{x} _{n , \vect{k}}$. Moreover, with the aid of the real-space probability distribution, one can further resolve the distinct behavior of bulk-edge mixed states, highlighted in cyan and green.

\section{Conclusions}

In this work, we developed a machine-learning-enabled tight-binding (TB) framework to overcome the fundamental limitations of simulating mesoscopic nanostructures. While density functional theory (DFT) provides accurate electronic structures, its prohibitive computational cost restricts simulations to systems with only a few hundred atoms per unit cell, far smaller than realistic nanostructures. Our strategy addresses this bottleneck by using Wannier tight-binding (WTB) parameters obtained from first-principles calculations of small nanostructures as training dataset for machine-learning (ML).

A key challenge in this approach is the gauge freedom of Wannier functions, which introduces arbitrariness in WTB parameters and obscures their dependence on nanostructure size and geometry, therefore hindering systematic parameter fitting and ML prediction. To resolve this, we constructed atomic-orbital-like gauge-independent (GI) bases and transformed the WTB model into a gauge-independent WTB (GI-WTB) model. This GI formulation restores clear geometric trends in the parameters, enabling robust fitting and ML interpolation across the geometric variable space and yielding the machine-learning GI-WTB (ML-GI-WTB) model capable of simulating nanostructures at realistic scales.

As a demonstration, we applied our machine-learning scheme to MoS$_{2}$ armchair-edge nanoribbons (A-NRs). The framework reproduced DFT band structures with high accuracy. Building on this agreement, we further used ML-GI-WTB model to predict parameter variations with respect to geometric variables and to simulate both energy band structures and wavefunctions for ribbons up to sub-$\mu$m widths.

The results show that the band gap of MoS$_{2}$ A-NRs rapidly saturates to a fixed value with increasing width. Beyond energy spectra, ML-GI-WTB provides complete real-space wavefunction information for all band states. Analysis of the relative average position at the $\Gamma$-point enables clear identification of bulk, edge, and bulk-edge mixed states with high spectral resolution.

In conclusion, ML-GI-WTB establishes a powerful and scalable methodology that combines first-principles-level reliability with computational efficiency. This framework enables predictive modeling of nanostructures at mesoscopic scales, provides a foundation for systematic studies of size- and geometry-dependent electronic properties, and offers significant potential for guiding the design of next-generation quantum devices.

\begin{acknowledgments}
This work is supported by the National Science and Technology Council (NSTC) of Taiwan under Grants NSTC 113-2112-M-A49-035-MY3, NSTC 114-2124-M-A49-004-, NSTC 114-2923-M-A49-005-MY3, and NSTC-DAAD 114-2927-I-A49-501-, as well as by the National Center for High-Performance Computing (NCHC) of Taiwan.
\end{acknowledgments}

\appendix

\newpage

\section{Demonstrations of Non-Transferability of Parameters}
\label{appendix_non_transfer}

\begin{figure}
	\centering
	\includegraphics[width=\textwidth]{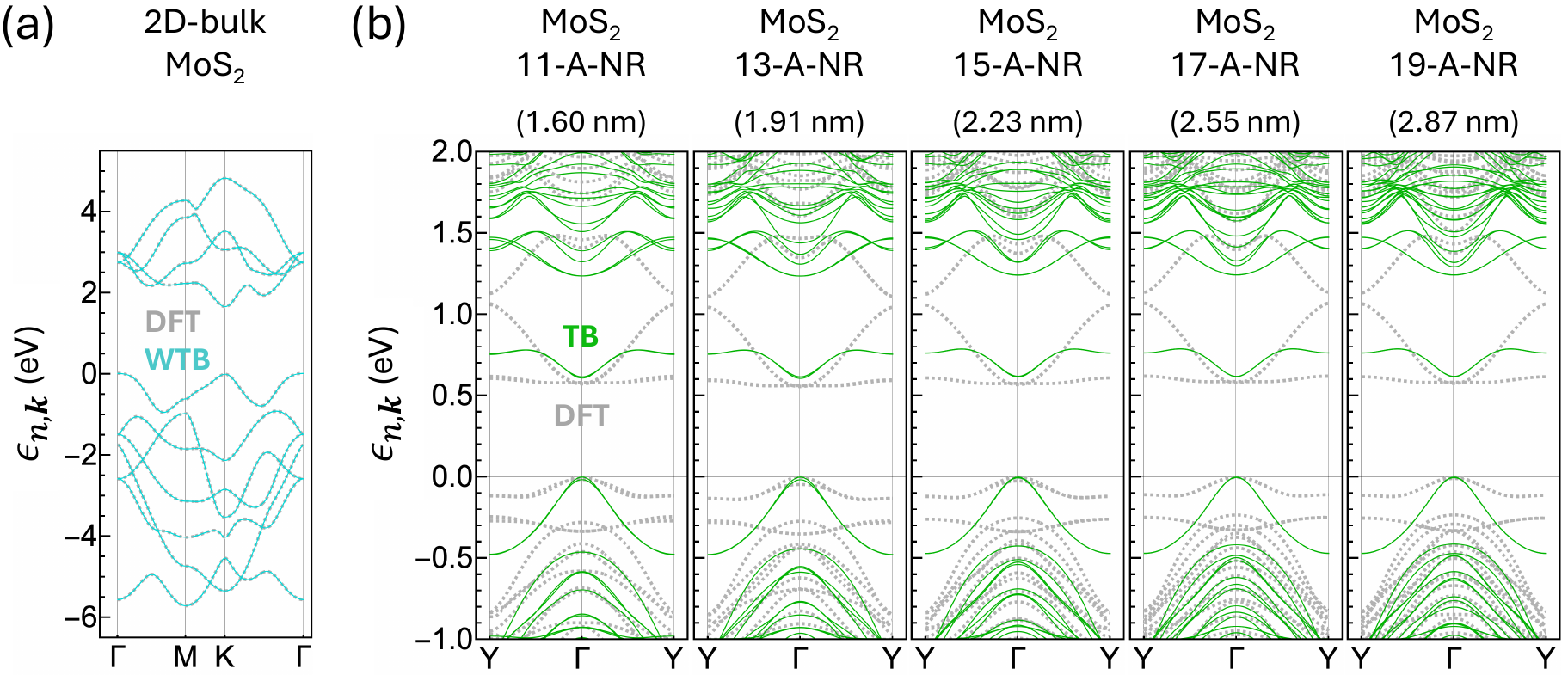}
	\caption{Band structure comparison between DFT (gray) and models. (a) WTB band structure (cyan) of 2D-bulk monolayer MoS$_{2}$. (b) Band structure of monolayer MoS$_{2}$ $N_a$-A-NRs from the TB model (green) constructed using 2D-bulk WTB parameters in (a).}
	\label{fig_non_transfer}
\end{figure}

To demonstrate the non-transferability of parameters, we use WTB parameters from 2D-bulk monolayer MoS$_{2}$ to construct TB models for monolayer MoS$_{2}$ $N _{a}$-A-NRs. 
Figure~\ref{fig_non_transfer}(a) shows the WTB band structure of 2D-bulk monolayer MoS$_{2}$, which agrees well with the DFT result. 
Using these 2D-bulk WTB parameters, we construct a new TB model for monolayer MoS$_{2}$ $N_{a}$-A-NRs. 
As shown in Fig.~\ref{fig_non_transfer}(b), the resulting band structure from this TB model deviates significantly from DFT, clearly demonstrating the non-transferability of bulk-derived parameters.

\section{Gauge-Independent Basis Set for Monolayer MoS$_{2}$ $N _{a}$-A-NRs}
\label{appendix_hybridized_basis}

As discussed in Sections~\ref{section_parameter_fitting} and \ref{section_GI_WTB_of_MoS2_NR}, the gauge-dependent basis set $\mathcal{S}^{\lambda}$ for $\lambda = N_{a}\text{-A-NR}$, generated through the Wannierization process, can be explicitly expressed as
\begin{align}
	\mathcal{S} ^{\lambda} 
	= \left\{ \big| W _{I \alpha , \vect{R}}  ^{\lambda} \big\rangle \, \Big| \, 1 \le I \le 3 N _{a} \, , \, \alpha \in \mathcal{A} _{\Theta ( I )} \, , \, \vect{R} \in \mathcal{WS} \right\}, 
\end{align}
where $\Theta ( I ) = \delta _{\text{mod} (I , 3) , 0} + 1$ is the atomic species function. For transition-metal atoms, $I$ will be a multiple of 3. For chalcogens, $I$ will be any number except the multiples of 3. When $I$ is a multiple of 3, the modulo function, $\text{mod}(I , 3)$, returns $0$, leading to $\Theta (I) = 2$. Otherwise, the modulo function will yield non-zero integers, leading to $\Theta (I) = 1$. 
The orbital set $\mathcal{A} _{1} = \{ p _{z} , p _{x} , p _{y} \}$ corresponds to the $p$-orbitals associated with chalcogen atoms. The orbital set $\mathcal{A} _{2} = \{ d _{z ^{2}} , d _{xz} , d _{yz} , d _{x ^{2} - y ^{2}} , d _{xy} \}$ corresponds to the $d$-orbitals associated with transition-metal atoms.
The set $\mathcal{WS} = \{ n _{2} \vect{a} _{2} \, \big| \, n _{2} \in \mathbb{Z} \, , \, - \frac{N _{2} - 1}{2} \le n _{2} \le \frac{N _{2} - 1}{2} \}$ defines the lattice vectors within the Wigner-Seitz supercell under periodic boundary conditions (PBCs), where $N _{2}$ is inherited from the $\vect{k}$-point sampling grid $N _{1} \times N _{2} \times N _{3}$ used in the DFT calculations. Since the PBCs employed in this work are independent of $\lambda$, the set $\mathcal{WS}$ is also independent of $\lambda$.

For MoS$_{2}$ nanoribbons with $\lambda = N_{a}$-A-NR, the corresponding gauge-independent (GI) basis set is defined as
\begin{align}\label{GI_basis_set}
	\mathcal{S} ^{\lambda , \text{GI}} =
	\mathcal{S} ^{\lambda , \text{GI}} _{\text{L-Edge}} \, \cup \, 
	\mathcal{S} ^{\lambda , \text{GI}} _{\text{R-Edge}} \, \cup \,	
	\mathcal{S} ^{\lambda , \text{GI}} _{\text{Bulk}} \, ,
\end{align}
where 
\begin{align}\label{basis_set_NR_L}
	\mathcal{S} ^{\lambda , \text{GI}} _{\text{L-Edge}} 
	= \left\{ \mathcal{T} _{\vect{\ell}} \big| W _{I \alpha , \vect{0}}  ^{\lambda _{0}} \big\rangle \, \Big| \, 1 \le I \le 3 N _{edge} \, , \, \alpha \in  \mathcal{A} _{\Theta ( I )} \, , \, \vect{\ell} = \left( \vect{R} + \vect{\tau} _{I} ^{\lambda} \right) - \vect{\tau} _{I} ^{\lambda _{0}} \, , \, \vect{R} \in \mathcal{WS} \right\}
\end{align}
represents the basis functions centered in the left edge region,
\begin{align}\label{basis_set_NR_R}
	\mathcal{S} ^{\lambda , \text{GI}} _{\text{R-Edge}} 
	= \left\{ \mathcal{T} _{\vect{\ell}} \big| W _{I \alpha , \vect{0}}  ^{\lambda _{0}} \big\rangle \, \Big| \, 3 N _{edge} < I \le 6 N _{edge} \, , \, \alpha \in  \mathcal{A} _{\Theta( I )} \, ,\, \vect{\ell} = \left( \vect{R} + \vect{\tau} _{I} ^{\lambda} \right) - \vect{\tau} _{I} ^{\lambda _{0}}  \, , \, \vect{R}  \in \mathcal{WS} \right\}
\end{align}
represents those centered in the right edge region, and 
\begin{align}\label{basis_set_NR_C}
	\mathcal{S} ^{\lambda , \text{GI}} _{\text{Bulk}} 
	= \left\{ \mathcal{T} _{\vect{\ell}} \big| W _{I \alpha , \vect{0}}  ^{\text{2D-bulk}} \big\rangle \, \Big| \, 1 \le I \le 3 \, , \, \alpha \in  \mathcal{A} _{\Theta ( I )} \, , \, \vect{\ell} = \left( \vect{R} + \vect{\tau} _{I ^{\prime}} ^{\lambda} \right) - \vect{\tau} _{I} ^{\text{2D-bulk}} \, , \, I ^{\prime} \in \mathcal{S} _{I} ^{\lambda} \, , \, \vect{R} \in \mathcal{WS} \right\}
\end{align}
represents those centered in the bulk region. 
The operator $\mathcal{T} _{\vect{\ell}}$ denotes the translation operator defined by $\mathcal{T} _{\vect{\ell}} \left| \vect{r} \right\rangle = \left| \vect{r} + \vect{\ell} \right\rangle$, where $\vect{\ell}$ is the corresponding displacement vector. The quantity $N _{edge}$ denotes the number of atomic chains comprising the left or right edge region, and $\lambda _{0} = N _{a} ^{0}\text{-A-NR}$ specifies the ribbon with designated width (see the discussion in Section~\ref{section_GI_WTB_of_MoS2_NR}). In Eq.~\eqref{basis_set_NR_C}, we define
\begin{align}\label{bulk_region_index_set}
	\mathcal{S} _{I} ^{\lambda} = \left\{ 6 N _{edge} + I + 3 \xi \, \big| \, \xi \in \mathbb{Z} \, , \, 6 N _{edge} < 6 N _{edge} + I + 3 \xi \le 3 N _{a} \right\} .
\end{align}
In this work, we have consistently used $N _{edge} = 4$ and $N _{a} ^{0} = 11$ throughout the analysis.  
The GI basis set defined in Eq.~\eqref{GI_basis_set} serves as the foundation for the discussions in Section~\ref{section_GI_WTB_of_MoS2_NR} and the sections that follow.

\section{GI-WTB Model Band Structures}
\label{appendix_GI_WTB_bands}

To further verify the reliability of the proposed GI-WTB model, we present additional band structures for MoS$_{2}$ $N _{a}$-A-NRs with $N _{a} = 13, 15, 17, 19$. In all cases, the results are consistent with DFT, confirming the robustness of our model.

\begin{figure}
	\centering
	\includegraphics[width=\textwidth]{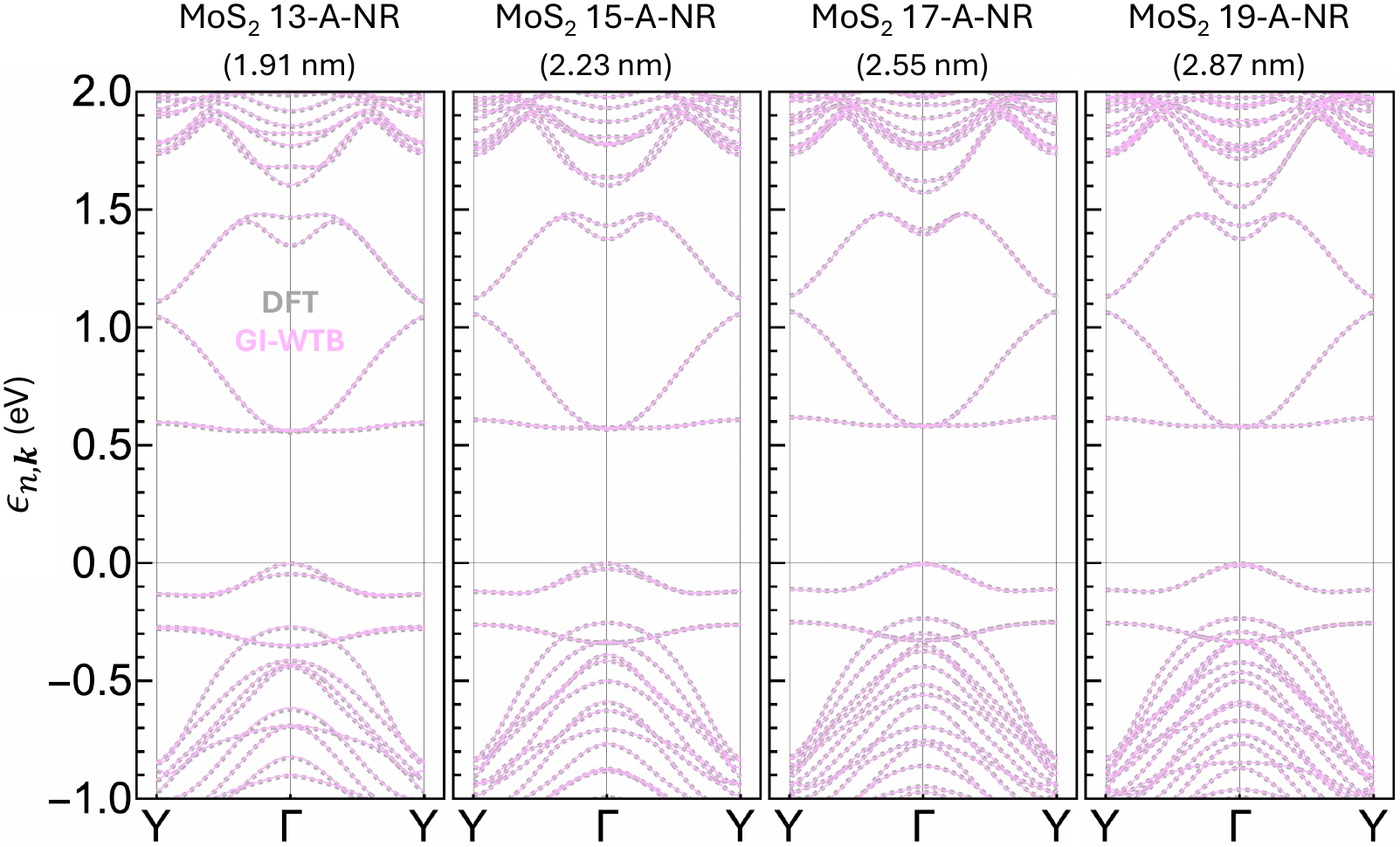}
	\caption{	Band structures of monolayer MoS$_{2}$ $N _{a}$-A-NRs with $N _{a} = 13, 15, 17, 19$, calculated using DFT (gray) and the GI-WTB model (pink).}
	\label{fig_GI_WTB_bands}
\end{figure}

\bibliography{ribbon_refs}

\begin{thebibliography}{14}%
\makeatletter
\providecommand \@ifxundefined [1]{%
 \@ifx{#1\undefined}
}%
\providecommand \@ifnum [1]{%
 \ifnum #1\expandafter \@firstoftwo
 \else \expandafter \@secondoftwo
 \fi
}%
\providecommand \@ifx [1]{%
 \ifx #1\expandafter \@firstoftwo
 \else \expandafter \@secondoftwo
 \fi
}%
\providecommand \natexlab [1]{#1}%
\providecommand \enquote  [1]{``#1''}%
\providecommand \bibnamefont  [1]{#1}%
\providecommand \bibfnamefont [1]{#1}%
\providecommand \citenamefont [1]{#1}%
\providecommand \href@noop [0]{\@secondoftwo}%
\providecommand \href [0]{\begingroup \@sanitize@url \@href}%
\providecommand \@href[1]{\@@startlink{#1}\@@href}%
\providecommand \@@href[1]{\endgroup#1\@@endlink}%
\providecommand \@sanitize@url [0]{\catcode `\\12\catcode `\$12\catcode
  `\&12\catcode `\#12\catcode `\^12\catcode `\_12\catcode `\%12\relax}%
\providecommand \@@startlink[1]{}%
\providecommand \@@endlink[0]{}%
\providecommand \url  [0]{\begingroup\@sanitize@url \@url }%
\providecommand \@url [1]{\endgroup\@href {#1}{\urlprefix }}%
\providecommand \urlprefix  [0]{URL }%
\providecommand \Eprint [0]{\href }%
\providecommand \doibase [0]{http://dx.doi.org/}%
\providecommand \selectlanguage [0]{\@gobble}%
\providecommand \bibinfo  [0]{\@secondoftwo}%
\providecommand \bibfield  [0]{\@secondoftwo}%
\providecommand \translation [1]{[#1]}%
\providecommand \BibitemOpen [0]{}%
\providecommand \bibitemStop [0]{}%
\providecommand \bibitemNoStop [0]{.\EOS\space}%
\providecommand \EOS [0]{\spacefactor3000\relax}%
\providecommand \BibitemShut  [1]{\csname bibitem#1\endcsname}%
\let\auto@bib@innerbib\@empty
%</preamble>
\bibitem [{\citenamefont {Kresse}\ and\ \citenamefont
  {Furthm\"uller}(1996)}]{kresse1996efficient}%
  \BibitemOpen
  \bibfield  {author} {\bibinfo {author} {\bibfnamefont {G.}~\bibnamefont
  {Kresse}}\ and\ \bibinfo {author} {\bibfnamefont {J.}~\bibnamefont
  {Furthm\"uller}},\ }\href {\doibase 10.1103/PhysRevB.54.11169} {\bibfield
  {journal} {\bibinfo  {journal} {Phys. Rev. B}\ }\textbf {\bibinfo {volume}
  {54}},\ \bibinfo {pages} {11169} (\bibinfo {year} {1996})}\BibitemShut
  {NoStop}%
\bibitem [{\citenamefont {Giannozzi}\ \emph {et~al.}(2017)\citenamefont
  {Giannozzi}, \citenamefont {Andreussi}, \citenamefont {Brumme}, \citenamefont
  {Bunau}, \citenamefont {Nardelli}, \citenamefont {Calandra}, \citenamefont
  {Car}, \citenamefont {Cavazzoni}, \citenamefont {Ceresoli}, \citenamefont
  {Cococcioni}, \citenamefont {Colonna}, \citenamefont {Carnimeo},
  \citenamefont {Corso}, \citenamefont {de~Gironcoli}, \citenamefont {Delugas},
  \citenamefont {Jr}, \citenamefont {Ferretti}, \citenamefont {Floris},
  \citenamefont {Fratesi}, \citenamefont {Fugallo}, \citenamefont {Gebauer},
  \citenamefont {Gerstmann}, \citenamefont {Giustino}, \citenamefont {Gorni},
  \citenamefont {Jia}, \citenamefont {Kawamura}, \citenamefont {Ko},
  \citenamefont {Kokalj}, \citenamefont {Küçükbenli}, \citenamefont
  {Lazzeri}, \citenamefont {Marsili}, \citenamefont {Marzari}, \citenamefont
  {Mauri}, \citenamefont {Nguyen}, \citenamefont {Nguyen}, \citenamefont {de-la
  Roza}, \citenamefont {Paulatto}, \citenamefont {Poncé}, \citenamefont
  {Rocca}, \citenamefont {Sabatini}, \citenamefont {Santra}, \citenamefont
  {Schlipf}, \citenamefont {Seitsonen}, \citenamefont {Smogunov}, \citenamefont
  {Timrov}, \citenamefont {Thonhauser}, \citenamefont {Umari}, \citenamefont
  {Vast}, \citenamefont {Wu},\ and\ \citenamefont {Baroni}}]{QE-2017}%
  \BibitemOpen
  \bibfield  {author} {\bibinfo {author} {\bibfnamefont {P.}~\bibnamefont
  {Giannozzi}}, \bibinfo {author} {\bibfnamefont {O.}~\bibnamefont
  {Andreussi}}, \bibinfo {author} {\bibfnamefont {T.}~\bibnamefont {Brumme}},
  \bibinfo {author} {\bibfnamefont {O.}~\bibnamefont {Bunau}}, \bibinfo
  {author} {\bibfnamefont {M.~B.}\ \bibnamefont {Nardelli}}, \bibinfo {author}
  {\bibfnamefont {M.}~\bibnamefont {Calandra}}, \bibinfo {author}
  {\bibfnamefont {R.}~\bibnamefont {Car}}, \bibinfo {author} {\bibfnamefont
  {C.}~\bibnamefont {Cavazzoni}}, \bibinfo {author} {\bibfnamefont
  {D.}~\bibnamefont {Ceresoli}}, \bibinfo {author} {\bibfnamefont
  {M.}~\bibnamefont {Cococcioni}}, \bibinfo {author} {\bibfnamefont
  {N.}~\bibnamefont {Colonna}}, \bibinfo {author} {\bibfnamefont
  {I.}~\bibnamefont {Carnimeo}}, \bibinfo {author} {\bibfnamefont {A.~D.}\
  \bibnamefont {Corso}}, \bibinfo {author} {\bibfnamefont {S.}~\bibnamefont
  {de~Gironcoli}}, \bibinfo {author} {\bibfnamefont {P.}~\bibnamefont
  {Delugas}}, \bibinfo {author} {\bibfnamefont {R.~A.~D.}\ \bibnamefont {Jr}},
  \bibinfo {author} {\bibfnamefont {A.}~\bibnamefont {Ferretti}}, \bibinfo
  {author} {\bibfnamefont {A.}~\bibnamefont {Floris}}, \bibinfo {author}
  {\bibfnamefont {G.}~\bibnamefont {Fratesi}}, \bibinfo {author} {\bibfnamefont
  {G.}~\bibnamefont {Fugallo}}, \bibinfo {author} {\bibfnamefont
  {R.}~\bibnamefont {Gebauer}}, \bibinfo {author} {\bibfnamefont
  {U.}~\bibnamefont {Gerstmann}}, \bibinfo {author} {\bibfnamefont
  {F.}~\bibnamefont {Giustino}}, \bibinfo {author} {\bibfnamefont
  {T.}~\bibnamefont {Gorni}}, \bibinfo {author} {\bibfnamefont
  {J.}~\bibnamefont {Jia}}, \bibinfo {author} {\bibfnamefont {M.}~\bibnamefont
  {Kawamura}}, \bibinfo {author} {\bibfnamefont {H.-Y.}\ \bibnamefont {Ko}},
  \bibinfo {author} {\bibfnamefont {A.}~\bibnamefont {Kokalj}}, \bibinfo
  {author} {\bibfnamefont {E.}~\bibnamefont {Küçükbenli}}, \bibinfo {author}
  {\bibfnamefont {M.}~\bibnamefont {Lazzeri}}, \bibinfo {author} {\bibfnamefont
  {M.}~\bibnamefont {Marsili}}, \bibinfo {author} {\bibfnamefont
  {N.}~\bibnamefont {Marzari}}, \bibinfo {author} {\bibfnamefont
  {F.}~\bibnamefont {Mauri}}, \bibinfo {author} {\bibfnamefont {N.~L.}\
  \bibnamefont {Nguyen}}, \bibinfo {author} {\bibfnamefont {H.-V.}\
  \bibnamefont {Nguyen}}, \bibinfo {author} {\bibfnamefont {A.~O.}\
  \bibnamefont {de-la Roza}}, \bibinfo {author} {\bibfnamefont
  {L.}~\bibnamefont {Paulatto}}, \bibinfo {author} {\bibfnamefont
  {S.}~\bibnamefont {Poncé}}, \bibinfo {author} {\bibfnamefont
  {D.}~\bibnamefont {Rocca}}, \bibinfo {author} {\bibfnamefont
  {R.}~\bibnamefont {Sabatini}}, \bibinfo {author} {\bibfnamefont
  {B.}~\bibnamefont {Santra}}, \bibinfo {author} {\bibfnamefont
  {M.}~\bibnamefont {Schlipf}}, \bibinfo {author} {\bibfnamefont {A.~P.}\
  \bibnamefont {Seitsonen}}, \bibinfo {author} {\bibfnamefont {A.}~\bibnamefont
  {Smogunov}}, \bibinfo {author} {\bibfnamefont {I.}~\bibnamefont {Timrov}},
  \bibinfo {author} {\bibfnamefont {T.}~\bibnamefont {Thonhauser}}, \bibinfo
  {author} {\bibfnamefont {P.}~\bibnamefont {Umari}}, \bibinfo {author}
  {\bibfnamefont {N.}~\bibnamefont {Vast}}, \bibinfo {author} {\bibfnamefont
  {X.}~\bibnamefont {Wu}}, \ and\ \bibinfo {author} {\bibfnamefont
  {S.}~\bibnamefont {Baroni}},\ }\href
  {http://stacks.iop.org/0953-8984/29/i=46/a=465901} {\bibfield  {journal}
  {\bibinfo  {journal} {J. Phys. Condens. Matter}\ }\textbf {\bibinfo {volume}
  {29}},\ \bibinfo {pages} {465901} (\bibinfo {year} {2017})}\BibitemShut
  {NoStop}%
\bibitem [{\citenamefont {Chen}\ \emph {et~al.}(2019)\citenamefont {Chen},
  \citenamefont {Kim}, \citenamefont {Chen}, \citenamefont {Yuan},
  \citenamefont {Bashir}, \citenamefont {Lou}, \citenamefont {van~der Zande},\
  and\ \citenamefont {King}}]{chen2019monolayer}%
  \BibitemOpen
  \bibfield  {author} {\bibinfo {author} {\bibfnamefont {S.}~\bibnamefont
  {Chen}}, \bibinfo {author} {\bibfnamefont {S.}~\bibnamefont {Kim}}, \bibinfo
  {author} {\bibfnamefont {W.}~\bibnamefont {Chen}}, \bibinfo {author}
  {\bibfnamefont {J.}~\bibnamefont {Yuan}}, \bibinfo {author} {\bibfnamefont
  {R.}~\bibnamefont {Bashir}}, \bibinfo {author} {\bibfnamefont
  {J.}~\bibnamefont {Lou}}, \bibinfo {author} {\bibfnamefont {A.~M.}\
  \bibnamefont {van~der Zande}}, \ and\ \bibinfo {author} {\bibfnamefont
  {W.~P.}\ \bibnamefont {King}},\ }\href {\doibase
  10.1021/acs.nanolett.9b00271} {\bibfield  {journal} {\bibinfo  {journal}
  {Nano Lett.}\ }\textbf {\bibinfo {volume} {19}},\ \bibinfo {pages} {2092}
  (\bibinfo {year} {2019})}\BibitemShut {NoStop}%
\bibitem [{\citenamefont {Peng}\ \emph {et~al.}(2019)\citenamefont {Peng},
  \citenamefont {Lo}, \citenamefont {Li}, \citenamefont {Huang}, \citenamefont
  {Chen}, \citenamefont {Lee}, \citenamefont {Yang},\ and\ \citenamefont
  {Cheng}}]{peng2019distinctive}%
  \BibitemOpen
  \bibfield  {author} {\bibinfo {author} {\bibfnamefont {G.-H.}\ \bibnamefont
  {Peng}}, \bibinfo {author} {\bibfnamefont {P.-Y.}\ \bibnamefont {Lo}},
  \bibinfo {author} {\bibfnamefont {W.-H.}\ \bibnamefont {Li}}, \bibinfo
  {author} {\bibfnamefont {Y.-C.}\ \bibnamefont {Huang}}, \bibinfo {author}
  {\bibfnamefont {Y.-H.}\ \bibnamefont {Chen}}, \bibinfo {author}
  {\bibfnamefont {C.-H.}\ \bibnamefont {Lee}}, \bibinfo {author} {\bibfnamefont
  {C.-K.}\ \bibnamefont {Yang}}, \ and\ \bibinfo {author} {\bibfnamefont
  {S.-J.}\ \bibnamefont {Cheng}},\ }\href {\doibase
  10.1021/acs.nanolett.8b04786} {\bibfield  {journal} {\bibinfo  {journal}
  {Nano Lett.}\ }\textbf {\bibinfo {volume} {19}},\ \bibinfo {pages} {2299}
  (\bibinfo {year} {2019})}\BibitemShut {NoStop}%
\bibitem [{\citenamefont {Lo}\ \emph {et~al.}(2021)\citenamefont {Lo},
  \citenamefont {Peng}, \citenamefont {Li}, \citenamefont {Yang},\ and\
  \citenamefont {Cheng}}]{lo2021fullzone}%
  \BibitemOpen
  \bibfield  {author} {\bibinfo {author} {\bibfnamefont {P.-Y.}\ \bibnamefont
  {Lo}}, \bibinfo {author} {\bibfnamefont {G.-H.}\ \bibnamefont {Peng}},
  \bibinfo {author} {\bibfnamefont {W.-H.}\ \bibnamefont {Li}}, \bibinfo
  {author} {\bibfnamefont {Y.}~\bibnamefont {Yang}}, \ and\ \bibinfo {author}
  {\bibfnamefont {S.-J.}\ \bibnamefont {Cheng}},\ }\href {\doibase
  10.1103/PhysRevResearch.3.043198} {\bibfield  {journal} {\bibinfo  {journal}
  {Phys. Rev. Research}\ }\textbf {\bibinfo {volume} {3}},\ \bibinfo {pages}
  {043198} (\bibinfo {year} {2021})}\BibitemShut {NoStop}%
\bibitem [{\citenamefont {Shih}\ \emph {et~al.}(2025)\citenamefont {Shih},
  \citenamefont {Peng}, \citenamefont {Lo}, \citenamefont {Li}, \citenamefont
  {Xu}, \citenamefont {Chien},\ and\ \citenamefont
  {Cheng}}]{shih2025signatures}%
  \BibitemOpen
  \bibfield  {author} {\bibinfo {author} {\bibfnamefont {C.-H.}\ \bibnamefont
  {Shih}}, \bibinfo {author} {\bibfnamefont {G.-H.}\ \bibnamefont {Peng}},
  \bibinfo {author} {\bibfnamefont {P.-Y.}\ \bibnamefont {Lo}}, \bibinfo
  {author} {\bibfnamefont {W.-H.}\ \bibnamefont {Li}}, \bibinfo {author}
  {\bibfnamefont {M.-L.}\ \bibnamefont {Xu}}, \bibinfo {author} {\bibfnamefont
  {C.-H.}\ \bibnamefont {Chien}}, \ and\ \bibinfo {author} {\bibfnamefont
  {S.-J.}\ \bibnamefont {Cheng}},\ }\href {\doibase 10.1103/qmyy-vn1x}
  {\bibfield  {journal} {\bibinfo  {journal} {Phys. Rev. B}\ }\textbf {\bibinfo
  {volume} {111}},\ \bibinfo {pages} {245422} (\bibinfo {year}
  {2025})}\BibitemShut {NoStop}%
\bibitem [{\citenamefont {Qiu}\ \emph {et~al.}(2015)\citenamefont {Qiu},
  \citenamefont {Cao},\ and\ \citenamefont {Louie}}]{qiu2015nonanalyticity}%
  \BibitemOpen
  \bibfield  {author} {\bibinfo {author} {\bibfnamefont {D.~Y.}\ \bibnamefont
  {Qiu}}, \bibinfo {author} {\bibfnamefont {T.}~\bibnamefont {Cao}}, \ and\
  \bibinfo {author} {\bibfnamefont {S.~G.}\ \bibnamefont {Louie}},\ }\href
  {\doibase 10.1103/PhysRevLett.115.176801} {\bibfield  {journal} {\bibinfo
  {journal} {Phys. Rev. Lett.}\ }\textbf {\bibinfo {volume} {115}},\ \bibinfo
  {pages} {176801} (\bibinfo {year} {2015})}\BibitemShut {NoStop}%
\bibitem [{\citenamefont {Deilmann}\ and\ \citenamefont
  {Thygesen}(2019)}]{Deilmann_2019}%
  \BibitemOpen
  \bibfield  {author} {\bibinfo {author} {\bibfnamefont {T.}~\bibnamefont
  {Deilmann}}\ and\ \bibinfo {author} {\bibfnamefont {K.~S.}\ \bibnamefont
  {Thygesen}},\ }\href {\doibase 10.1088/2053-1583/ab0e1d} {\bibfield
  {journal} {\bibinfo  {journal} {2D Mater.}\ }\textbf {\bibinfo {volume}
  {6}},\ \bibinfo {pages} {035003} (\bibinfo {year} {2019})}\BibitemShut
  {NoStop}%
\bibitem [{\citenamefont {Marzari}\ and\ \citenamefont
  {Vanderbilt}(1997)}]{PhysRevB.56.12847}%
  \BibitemOpen
  \bibfield  {author} {\bibinfo {author} {\bibfnamefont {N.}~\bibnamefont
  {Marzari}}\ and\ \bibinfo {author} {\bibfnamefont {D.}~\bibnamefont
  {Vanderbilt}},\ }\href {\doibase 10.1103/PhysRevB.56.12847} {\bibfield
  {journal} {\bibinfo  {journal} {Phys. Rev. B}\ }\textbf {\bibinfo {volume}
  {56}},\ \bibinfo {pages} {12847} (\bibinfo {year} {1997})}\BibitemShut
  {NoStop}%
\bibitem [{\citenamefont {Mostofi}\ \emph {et~al.}(2014)\citenamefont
  {Mostofi}, \citenamefont {Yates}, \citenamefont {Pizzi}, \citenamefont {Lee},
  \citenamefont {Souza}, \citenamefont {Vanderbilt},\ and\ \citenamefont
  {Marzari}}]{mostofi2014an}%
  \BibitemOpen
  \bibfield  {author} {\bibinfo {author} {\bibfnamefont {A.~A.}\ \bibnamefont
  {Mostofi}}, \bibinfo {author} {\bibfnamefont {J.~R.}\ \bibnamefont {Yates}},
  \bibinfo {author} {\bibfnamefont {G.}~\bibnamefont {Pizzi}}, \bibinfo
  {author} {\bibfnamefont {Y.-S.}\ \bibnamefont {Lee}}, \bibinfo {author}
  {\bibfnamefont {I.}~\bibnamefont {Souza}}, \bibinfo {author} {\bibfnamefont
  {D.}~\bibnamefont {Vanderbilt}}, \ and\ \bibinfo {author} {\bibfnamefont
  {N.}~\bibnamefont {Marzari}},\ }\href {\doibase
  https://doi.org/10.1016/j.cpc.2014.05.003} {\bibfield  {journal} {\bibinfo
  {journal} {Comput. Phys. Commun.}\ }\textbf {\bibinfo {volume} {185}},\
  \bibinfo {pages} {2309} (\bibinfo {year} {2014})}\BibitemShut {NoStop}%
\bibitem [{\citenamefont {Ridolfi}\ \emph {et~al.}(2017)\citenamefont
  {Ridolfi}, \citenamefont {Lima}, \citenamefont {Mucciolo},\ and\
  \citenamefont {Lewenkopf}}]{PhysRevB.95.035430}%
  \BibitemOpen
  \bibfield  {author} {\bibinfo {author} {\bibfnamefont {E.}~\bibnamefont
  {Ridolfi}}, \bibinfo {author} {\bibfnamefont {L.~R.~F.}\ \bibnamefont
  {Lima}}, \bibinfo {author} {\bibfnamefont {E.~R.}\ \bibnamefont {Mucciolo}},
  \ and\ \bibinfo {author} {\bibfnamefont {C.~H.}\ \bibnamefont {Lewenkopf}},\
  }\href {\doibase 10.1103/PhysRevB.95.035430} {\bibfield  {journal} {\bibinfo
  {journal} {Phys. Rev. B}\ }\textbf {\bibinfo {volume} {95}},\ \bibinfo
  {pages} {035430} (\bibinfo {year} {2017})}\BibitemShut {NoStop}%
\bibitem [{\citenamefont {Have}\ \emph {et~al.}(2019)\citenamefont {Have},
  \citenamefont {Peres},\ and\ \citenamefont {Pedersen}}]{PhysRevB.100.045411}%
  \BibitemOpen
  \bibfield  {author} {\bibinfo {author} {\bibfnamefont {J.}~\bibnamefont
  {Have}}, \bibinfo {author} {\bibfnamefont {N.~M.~R.}\ \bibnamefont {Peres}},
  \ and\ \bibinfo {author} {\bibfnamefont {T.~G.}\ \bibnamefont {Pedersen}},\
  }\href {\doibase 10.1103/PhysRevB.100.045411} {\bibfield  {journal} {\bibinfo
   {journal} {Phys. Rev. B}\ }\textbf {\bibinfo {volume} {100}},\ \bibinfo
  {pages} {045411} (\bibinfo {year} {2019})}\BibitemShut {NoStop}%
\bibitem [{\citenamefont {Gibertini}\ and\ \citenamefont
  {Marzari}(2015)}]{gibertini2015emergence}%
  \BibitemOpen
  \bibfield  {author} {\bibinfo {author} {\bibfnamefont {M.}~\bibnamefont
  {Gibertini}}\ and\ \bibinfo {author} {\bibfnamefont {N.}~\bibnamefont
  {Marzari}},\ }\href {\doibase 10.1021/acs.nanolett.5b02834} {\bibfield
  {journal} {\bibinfo  {journal} {Nano Lett.}\ }\textbf {\bibinfo {volume}
  {15}},\ \bibinfo {pages} {6229} (\bibinfo {year} {2015})}\BibitemShut
  {NoStop}%
\bibitem [{\citenamefont {Perdew}\ \emph {et~al.}(1996)\citenamefont {Perdew},
  \citenamefont {Burke},\ and\ \citenamefont
  {Ernzerhof}}]{perdew1996generalized}%
  \BibitemOpen
  \bibfield  {author} {\bibinfo {author} {\bibfnamefont {J.~P.}\ \bibnamefont
  {Perdew}}, \bibinfo {author} {\bibfnamefont {K.}~\bibnamefont {Burke}}, \
  and\ \bibinfo {author} {\bibfnamefont {M.}~\bibnamefont {Ernzerhof}},\ }\href
  {\doibase 10.1103/PhysRevLett.77.3865} {\bibfield  {journal} {\bibinfo
  {journal} {Phys. Rev. Lett.}\ }\textbf {\bibinfo {volume} {77}},\ \bibinfo
  {pages} {3865} (\bibinfo {year} {1996})}\BibitemShut {NoStop}%
\end{thebibliography}%

\end{document}